\documentclass{article}

\usepackage[english]{babel}
\usepackage[letterpaper,top=2cm,bottom=2cm,left=2cm,right=2cm,marginparwidth=1.5cm]{geometry}
\usepackage{caption}
\usepackage{subcaption}
\usepackage{placeins}
\usepackage{amsmath}
\usepackage{natbib}
\usepackage{graphicx}
\usepackage{multirow}
\usepackage{xcolor}
\usepackage{makecell}
\usepackage{esint}
\usepackage{authblk}
\usepackage{siunitx}
\usepackage[colorlinks=true, allcolors=blue]{hyperref}

\title{Magnetic flutter effect on validated edge turbulence simulations}
\author[]{Kaiyu Zhang\thanks{corresponding author: kaiyu.zhang@ipp.mpg.de}}
\author[]{Wladimir Zholobenko}
\author[]{Andreas Stegmeir}
\author[]{Konrad Eder}
\author[]{Frank Jenko}
\affil{Max Planck Institute for Plasma Physics, Boltzmannstr. 2, 85748 Garching, Germany}

\date{}
\begin{document}
\maketitle

\begin{abstract}
Small magnetic fluctuations ($B_1/B_0 \sim 10^{-4}$) are intrinsically present in a magnetic confinement plasma due to turbulent currents.
While the perpendicular transport of particles and heat is typically dominated by fluctuations of the electric field, the parallel stream of plasma is affected by fluttering magnetic field lines. In particular through electrons, this indirectly impacts the turbulence dynamics.
Even in low beta conditions, we find that $E \times B$ turbulent transport can be reduced by more than a factor 2 when magnetic flutter is included in our validated edge turbulence simulations of L-mode ASDEX Upgrade. 
The primary reason for this is the stabilization of drift-Alfv\'en-waves, which reduces the phase shifts of density and temperature fluctuations with respect to potential fluctuations. 
This stabilization can be qualitatively explained by linear analytical theory, and appreciably reinforced by the flutter nonlinearity.
As a secondary effect, the steeper temperature gradients and thus higher $\eta_i$ increase the impact of the ion-temperature-gradient mode on overall turbulent transport.
With increasing beta, the stabilizing effect on $E\times B$ turbulence increases, balancing the destabilization by induction, until direct electromagnetic perpendicular transport is triggered. 
We conclude that including flutter is crucial for predictive edge turbulence simulations.
\end{abstract}

\section{Introduction}
In magnetic confinement devices, the magnetic field can always be perturbed by small magnetic fluctuations due to turbulent currents. Specifically, this occurs due to the magnetic induction in Ohm's law. This linear induction term is critical for the numerical performance of turbulence codes \citep{scott1997three,DANNERT200467,Dudson_2021,STEGMEIR2023108801} because it limits the time step by the shear Alfv\'en wave speed, which is still much slower than the diffusive time step restrictions in electrostatic models \citep{Dudson_2021}. 
A second electromagnetic effect is magnetic flutter. It arises as the plasma streams along the perturbed magnetic field lines. In contrast to induction, the role of flutter is more intricate as it permeates the entire turbulence system by introducing nonlinearity to each parallel operator. Consequently, the impact of preserving the flutter effect is not entirely clear and warrants further investigation.

Previous research indicates that the flutter effect on turbulent transport manifests in two distinct ways. On the one hand, magnetic flutter can trigger electromagnetic transport. In high-beta regimes, \citet{callen1977drift} first found that magnetic flutter enhances radial electron heat transport by disrupting the magnetic flux surface and forming magnetic islands. This phenomenon, referred to as `magnetic flutter transport,' stands out due to the rapid parallel transport entering the radial transport via the radial component of magnetic perturbations. Gyrokinetic simulations later confirmed that this flutter transport dominates electron heat transport in micro-tearing mode (MTM) turbulence \citep{doerk2011gyrokinetic}.
Conversely, in low-beta regimes, the significance of flutter transport is less obvious. 
Experimental measurements found that the magnetic fluctuations were too small to generate appreciable flutter transport \citep{mantica1991broadband,hidalgo1995edge}. 
The fluid simulations for drift-Alfv\'en-wave (DAW) turbulence demonstrated that the flutter transport was virtually negligible in comparison to $E \times B$ transport \citep{scott1997three,naulin2003electromagnetic}.
\citet{giacomin2022turbulent} found the flutter transport played a minor role in resistive ballooning regimes.
Regarding ion temperature gradient (ITG)-dominated turbulence in the tokamak core, the flutter transport remains inconsequential over a large $\beta$ range but only becomes prominent upon crossing the kinetic ballooning mode (KBM) threshold \citep{pueschel2010transport}.

On the other hand, magnetic flutter can indirectly impact electrostatic ($E \times B$) transport. This indirect influence was first investigated by \citet{jenko1999numerical}, who compared kinetic simulations with and without flutter for weakly collisional drift-Alfv\'en-wave (DAW) turbulence in a sheared slab geometry. Their study revealed that while flutter transport was nearly negligible, the electrostatic particle transport was significantly reduced by magnetic flutter. 
% Consequently, \citet{jenko1999numerical} argued that despite electrostatic transport dominance, the intrinsic dynamics of the turbulence remain electromagnetic due to the presence of magnetic flutter.
\citet{scott2001low} developed a linear analytical theory for the DAW instability and observed that the role of magnetic flutter is to stabilize linear perturbations and decrease the phase shift between perturbations of potential and density.
Similarly, electrostatic heat transport in ion-temperature-gradient (ITG) dominated turbulence was also found to be reduced by the electromagnetic effect \citep{snyder2001electromagnetic}. This phenomenon is related to the magnetic stabilization effect on ITG, which has been extensively studied \citep{tang1978microinstability,dong1987finite,kim1993electromagnetic,whelan2018nonlinear,hirose2000finite,citrin2014electromagnetic}.
In addition to dedicated models for DAW and ITG, recent comprehensive simulations for L-mode tokamak edge and scrape-off layer (SOL) regions have also confirmed the reduced transport. \citet{mandell2020electromagnetic} compared gyrokinetic simulations in both electrostatic and electromagnetic cases for a helical open-field-line box with Gkeyll. They observed that the temperatures and density profiles become shallower with electromagnetic effect in the SOL due to the reduced particle transport, although the individual contributions of magnetic induction and magnetic flutter were not distinguished in their study.

Despite the extensive research summarized above, two questions persist regarding the impact of magnetic flutter on edge turbulence, particularly in low-beta simulations. 
The first question is whether we can ignore flutter.
The urgency to address the question arises as edge turbulence codes strive for greater predictability. 
The theoretical studies view tokamak edge turbulence as consistently electromagnetic \citep{jenko2001nonlinear,kobayashi2020physics,eich2021separatrix}. 
% due to a large safety factor $q$ and a significant inverse scale length $L_\perp$. 
% Even in scenarios with lower beta values, the presence of a large safety factor $q$ or a significant inverse scale length $L_\perp$ at the tokamak edge suggests that flutter can still exert an influence \citep{jenko1999numerical,jenko2001nonlinear}.
However, from the perspective of code development, many comprehensive global full-size simulations in realistic diverted geometry choose to disregard magnetic flutter due to the small flutter transport \citep{stegmeir2019global,giacomin2020investigation,giacomin2021theory,zholobenko2021electric}. 
Despite these contrasting viewpoints, no quantitative report has been provided to definitively settle the argument and elucidate the differences in full-size edge turbulence simulations between models with and without flutter.

The second question pertains to the micro-instability through which magnetic flutter impacts edge turbulence. As previously mentioned, flutter can influence several micro-instabilities, including drift-Alfv\'en-wave (DAW), ion temperature gradient (ITG), micro-tearing mode (MTM), and kinetic ballooning mode (KBM). The extensive literature on these specific modes is readily available.
However, the complexity of edge turbulence in real reactors poses a challenge, as there is no dominant mode or instability \citep{scott2007tokamak}. 
Consequently, it remains unclear which modes flutter will affect in a real reactor setting. 
Addressing this question requires validated simulations that accurately represent the conditions in real reactors.
% It is obvious that magnetic flutter plays a vital role in high-beta edge turbulence due to the noticeable electromagnetic transport.
% But it remains a dispute whether we can neglect magnetic flutter in low-beta cases in terms of predicable turbulence simulations. 
% Some global kinetic electromagnetic simulations \citep{michels2022full,hager2022electromagnetic} are always performed with magnetic flutter intrinsically.
% Although the transport directly caused by magnetic flutter is frequently discussed by those electromagnetic simulations, the indirect influence of magnetic flutter has never been distinguished by a comparison study.
% In this context, to what extend magnetic flutter will alter the low-beta turbulence on edge and SOL in full-size realistic diverted geometry is not well understood yet.
% Magnetic flutter is particularly pronounced on tokamak edge and scrape-off Layer (SOL) \citep{jenko2001nonlinear}.
% Even without a large beta, the electromagnetic turbulence can play a role because of a large safety factor $q$ or a large inverse scale length $L_\perp$ at the tokamak edge \citep{kobayashi2020physics},
% Tokamak edge turbulence is almost always in the electromagnetic regime \citep{eich2021separatrix}.
 
We are motivated by the aforementioned questions to conduct an investigation into the comprehensive effects of magnetic flutter on edge turbulence through numerical simulations. Numerical modeling for the tokamak edge and SOL presents significant challenges because of the complex diverted geometry \citep{stegmeir2016field,chang2017gyrokinetic}, non-local turbulence spreading \citep{manz2015origin,manz2020diffusion}, and large turbulent fluctuations \citep{ritz1989fluctuation,zweben2007edge,garcia2007collisionality}.
To address these challenges, we have developed GRILLIX, a global `full-$f$' fluid turbulence code that employs the flux-coordinate independent (FCI) approach \citep{stegmeir2016field,stegmeir2017advances} to deal with the complex diverted geometries \citep{Body2019}. 
The comparable codes are BOUT++ \citep{zhu2021drift,walkden2022physics}, GBS \citep{paruta2019blob,giacomin2022gbs}, and SOLEDGE3X \citep{bufferand2021progress}, as well as gyrokinetic codes like GENE-X \citep{michels2021gene} and XGC \citep{hager2022electromagnetic}.
The implementation of magnetic flutter into GRILLIX has been verified using the method of manufactured solutions \citep{Salari_MMS_2000}. 
Notably, our simulations have been conducted for ASDEX Upgrade without any down-scaling, marking the first study to investigate the magnetic flutter effect on validated full-size edge turbulence simulations. 
Our findings highlight the crucial role of including magnetic flutter in predictive simulations, as it leads to a significant reduction in turbulent transport.

% due to the following reasons:
% (1) Global modeling is crucial because turbulence spreads throughout the edge and SOL regions \citep{manz2015origin,manz2020diffusion}, leading to non-local fluctuations.
% (2) The diverted geometry introduces major complexities to global models, especially at the separatrix and X-point, where field-aligned coordinates become singular .
% (3) A full-\textit{f} (full-f distribution function) treatment is necessary as fluctuations' amplitude (except for magnetic fluctuations) in the edge and SOL can be comparable to the plasma background \citep{ritz1989fluctuation,zweben2007edge,garcia2007collisionality}.
% (4) Despite using full-\textit{f} plasma fields, the magnetic fluctuation used for calculating magnetic flutter should remain as a delta-\textit{f} (delta-f distribution function) quantity to avoid double counting the Shafranov shift \citep{scott2006edge}. Hence, it becomes necessary to accurately extract pure fluctuations from the full-\textit{f} plasma turbulence throughout the entire simulation \citep{hager2022electromagnetic,halpern2016gbs}, which is a challenging task in physics and might also be computationally expensive \citep{hager2020gyrokinetic}. 

This paper is organized as follows:
Chapter \ref{sec:model} provides a detailed description of the numerical model, including the treatment of magnetic flutter and the simulation setup.
Chapter \ref{sec:results} presents and compares the simulation results between models with and without flutter.
In Chapter \ref{sec:discussion}, we delve into the mechanism of how magnetic flutter influences turbulence. This will be achieved through the combination of an analytical theory and numerical experiments.

\section{Model} \label{sec:model}
\subsection{Global drift reduced Braginskii equations}\label{sec:equations}
The drift-reduced Braginskii equations \citep{braginskii1965transport,zeiler1997nonlinear} are employed in this study.
The turbulence is strongly anisotropic, with a small perpendicular scale approximately equal to the ion Larmor radius $\rho_{s0}$, i.e., $L_\perp \sim \rho_{s0}$, but a large parallel scale approximately equal to the major radius $\sim R_0$ of the device, i.e.,  $L_\parallel \sim R_0$. 
The anisotropy yields the drift ordering parameter $\delta = R_0/\rho_{s0}$, which is consistently large in magnetically confined plasmas ($\delta \approx 3\times10^3$ in this work).
The velocity vectors are separated into parallel components ($u_\parallel$ for ions and $v_\parallel$ for electrons) and perpendicular drifts.
% components containing the $E\times B$ drift and diamagnetic drift. The polarisation drift velocity of ions is considered only in the vorticity (or quasi-neutrality) equation, while the $\nabla B$ drift is entirely ignored.

In terms of normalization, the reference temperatures and density are denoted as $T_0$ and $n_0$, respectively. The reference magnetic field $B_0$ represents the toroidal magnetic field strength on the axis. Then the nominal sound speed, Larmor radius, and the dynamical plasma beta are obtained as $c_{s0}=\sqrt{T_{0}/M_i}$, $\rho_{s0}=c\sqrt{T_0 M_i}/(e B_0)$, and $\beta_{0}=4\pi n_0 T_0 /B_0^2$, respectively, where $M_i$ is the ion mass, $c$ is the speed of light, and $e$ is the elementary charge. The time $t$ is normalized against $\tau_0=R_0/c_{s0}$.
The perpendicular scales $L_\perp$ are normalized against $\rho_{s0}$ and the parallel scales $L_\parallel$ against $R_0$ according to the assumption of strong anisotropy. 
The parallel ion velocity $u_\parallel$ and electron velocity $v_\parallel$ are normalized against $c_{s0}$, and the parallel current $j_\parallel$ against $e n_0 c_{s0}$. 
The electrostatic potential $\phi$ is normalized against $T_0/e$, while the parallel component of magnetic potential $A_\parallel$ is normalized against $\beta_0 B_0 \rho_{s0}$, resulting in the natural normalization of magnetic perturbation $B_1$ as $\beta_0 B_0$.
The normalized set of equations reads:
\begin{equation}
    \frac{\mathrm{d}}{\mathrm{d} t} n 
    = n \mathcal{C}(\phi)-\mathcal{C}\left(p_{e}\right)+{ \nabla \cdot(j_{\|} \mathbf{b})} - { \nabla \cdot(n u_{\|} \mathbf{b})}+ \mathcal{D}_n(n)+S_n\,,
    \label{eqn:6-field:cont}
\end{equation}

\begin{equation}
\nabla \cdot\left[\frac{n}{B^2}\left(\frac{\mathrm{d}}{\mathrm{d} t}+u_{\|} { \mathbf{b}\cdot\nabla}\right)\left(\nabla_{\perp} \phi+\frac{\nabla_{\perp} p_{i}}{n}\right)\right]
=-\mathcal{C}\left(p_{e}+ p_{i}\right)+{ \nabla \cdot(j_{\|} \mathbf{b})} -\frac{1}{6} \mathcal{C}(G)+\mathcal{D}_{\Omega}(\Omega)\,,
    \label{eqn:6-field:vort}
\end{equation}

\begin{equation}
    \left(\frac{\mathrm{d}}{\mathrm{d} t}+u_{\|} { \mathbf{b}\cdot\nabla}\right) u_{\|} 
    =-\frac{\mathbf{b}\cdot\nabla p_{e}}{n}-\frac{\mathbf{b}\cdot\nabla(p_{i})}{n}+T_{i} \mathcal{C}\left(u_{\|}\right)-\frac{2}{3} \frac{B^{3 / 2}}{n} { \mathbf{b}\cdot\nabla} \frac{G}{B^{3 / 2}}+\mathcal{D}_u\left(u_{\|}\right)\,,
    \label{eqn:6-field:upar}
\end{equation}

\begin{equation}
    \beta_0 \frac{\partial}{\partial t} A_{\|}+\mu\left(\frac{\mathrm{d}}{\mathrm{d} t}+v_{\|} { \mathbf{b}\cdot\nabla}\right) \frac{j_{\|}}{n}
    =-\left(\frac{\eta_{\| 0}}{T_{e}^{3 / 2}}\right) j_{\|}-{\mathbf{b}\cdot\nabla} \phi+\frac{\mathbf{b}\cdot\nabla p_{e}}{n}+0.71 {\mathbf{b}\cdot\nabla} T_{e}+\mathcal{D}_{\Psi}\left(\Psi_m\right)\,,
    \label{eqn:6-field:apar}
\end{equation}

\begin{align}
    \frac{3}{2}\left(\frac{\mathrm{d}}{\mathrm{d} t}+v_{\|} {\mathbf{b}\cdot\nabla}\right) T_{e}&
    =T_{e} \mathcal{C}(\phi)-\frac{T_{e}}{n} \mathcal{C}\left(nT_e\right)-\frac{5}{2} T_{e} \mathcal{C}\left(T_{e}\right)-T_{e} {\nabla\cdot(v_{\|}\mathbf{b})}  +0.71 \frac{T_{e}}{n} {\nabla\cdot(j_{\|}\mathbf{b})} \nonumber\\
    &+\frac{1}{n} {\nabla\cdot(q_{\parallel,e}\mathbf{b})} 
    -3 \nu_{e 0} \mu\left(\frac{n}{T_{e}^{3 / 2}}\right)\left(T_{e}- T_{i}\right)+\left(\frac{\eta_{\| 0}}{T_{e}^{3 / 2}}\right) \frac{j_{\|}^2}{n}+\frac{3}{2}\left(\mathcal{D}_{T_{e}}\left(T_{e}\right)+S_{T_{e}}\right)\,,
    \label{eqn:6-field:te}    
\end{align}

\begin{align}
\frac{3}{2}\left(\frac{\mathrm{d}}{\mathrm{d} t}+u_{\|} { \mathbf{b}\cdot\nabla}\right) T_{i}&=T_{i} \mathcal{C}(\phi)-\frac{T_{i}}{n} \mathcal{C}\left(nT_e\right)+\frac{5}{2} T_{i} \mathcal{C}\left(T_{i}\right)-T_{i} {\nabla\cdot(u_\parallel\mathbf{b})}+\frac{T_{i}}{n}{\nabla\cdot(j_\parallel\mathbf{b})} \nonumber\\
& +\frac{1}{n}  {\nabla\cdot(q_{\parallel,i}\mathbf{b})}
+3 \nu_{e 0} \mu\left(\frac{n}{T_{e}^{3 / 2}}\right)\left( T_{e}-T_{i}\right)+\frac{1}{3 \eta_{i 0}} \frac{G^2}{n T_{i}^{5 / 2}}+\frac{3}{2}\left(\mathcal{D}_{T_{i}}\left(T_{i}\right)+S_{T_{i}}\right)\,,
    \label{eqn:6-field:ti}    
\end{align}

\begin{equation}
\nabla_{\perp}^2 A_{\|}=-j_{\|}\,.
    \label{eqn:6-field:amper} 
\end{equation}
The equations \eqref{eqn:6-field:cont}-\eqref{eqn:6-field:amper} are continuity equation, vorticity equation, parallel momentum equation, Ohm's law, electron temperature equation, ion temperature equation and Ampere's law.
The curvature operator is $\mathcal{C}(f)\approx-2 \partial  f/\partial Z$.
The full derivative is defined as  $\mathrm{d}/\mathrm{d}t=\partial/\partial t + \mathbf{v}_E\cdot\nabla$ with $\mathbf{v}_E= (\mathbf{b}_0 \times \nabla \phi)/B_\mathrm{tor}$ being the $E\times B$ drift velocity.
$p_{e,i}=nT_{e,i}$ is the pressure.
$\mu={M_e}/{M_i}$ is the mass ratio of electron to ion.
$\nu_{e 0}$ is the dimensionless electron collisionality. $\eta_{\parallel0}=0.51\mu\nu_{e 0}$ is the dimensionless parallel resistivity. 
\begin{equation}\label{eqn:6-field:G}
G=-\eta_{i 0} T_i^{5/2}\left[\frac{2}{B^{3 / 2}} { \nabla}\cdot(u_{\|} B^{3 / 2}\mathbf{b})-\frac{1}{2}\left(\mathcal{C}(\phi)+\frac{1}{n} \mathcal{C}\left(p_{i}\right)\right)\right]\,,
\end{equation}
is the viscous function entering the equations for vorticity, parallel momentum and ion temperature.
% The Braginskii ion viscosity coefficient $\eta_{i0}$ in $G$ is only valid in the Pfirsch-Schluter regime. 
% However, in the confined region, tokamaks typically operate in the plateau regime in the plasma edge, and in the banana regime deeper inside, while the Pfirsch-Schluter (PS) regime is only reached in the collisional SOL.
$\eta_{i 0}$ is the neoclassical correction to the Braginskii ion viscosity $\eta_{i 0}=\eta_{i 0}^\mathrm{BG}/(1 + \nu_*^{-1})$ in the plateau regime, which is usually valid in the edge. $\nu_*$ is the collisionality parameter according to \cite{Hirshman1981}.
%where \cite{Rozhansky2009} and \cite{helander2005collisional}.
% \begin{equation}
%     \tilde{\eta}^\mathrm{i}_0 = \frac{\eta^i_0}{(1 + 0.3^{-3/2}\nu_*^{-1})(1 + \nu_*^{-1})},
%     \label{eq:neo}
% \end{equation}
% where 0.3 is the the inverse aspect ratio for ASDEX Upgrade, and $\nu_*$ is the collisionality parameter according to \cite{Hirshman1981}.
To avoid the inappropriately high Braginskii heat conductivities at low collisionality, a harmonic average between Braginskii heat flux $q^\mathrm{BG}_\parallel=-\chi^{e,i}_\parallel \nabla_\parallel T_{e,i}$ and the free streaming heat flux $q^\mathrm{FS}_\parallel=n T_{e,i}^{3/2}$ is applied \citep{zholobenko2021role}, 
\begin{equation}
    q^{e,i}_\parallel =
    \left(\frac{1}{q^\mathrm{BG}_\parallel}+\frac{1}{\alpha^\mathrm{FS}_{e,i}q^\mathrm{FS}_\parallel}
    \right)^{-1},
    \label{eq:q}
\end{equation}
with free streaming coefficients
$\alpha^\mathrm{FS}_{e,i}$ as free parameters varying from 0.03 to 3.0 \citep{fundamenski2005parallel}. Nonetheless, particularly for the electron heat conduction, a 3D linear solver is used for its implicit treatment \citep{Zholobenko2019}, to allow for a larger time step limited only by shear Alfv\'en waves.
% \begin{equation}
%     \tilde{\chi}^{e,i}_\parallel = \chi^{e,i}_\parallel
%     \left(1+\frac{\chi^{e,i}_\parallel}{\alpha^\mathrm{FS}_{e,i}+}
%     \right)^{-1},
%     \label{eq:neo}
% \end{equation}

\subsection{Treatment of magnetic flutter}\label{sec:flutter}
The parallel field operators $\mathbf{b} \cdot \nabla \circ$ and $\nabla \cdot (\circ \mathbf{b})$ involve the unit vector of the magnetic field $\mathbf{b}$, which contains the magnetic equilibrium $\mathbf{b}_0$ and magnetic perturbation $\mathbf{b}_1$.
The discretization of the equilibrium operators $\mathbf{b}_0 \cdot \nabla \circ$ and $\nabla \cdot (\circ \mathbf{b}_0)$ is done by field line tracing within the framework of FCI \citep{stegmeir2016field,stegmeir2017advances}.

The flutter operators are linked to the magnetic perturbation $\mathbf{b}_1$. 
We assume that $\mathbf{b}_1$ is perpendicular to the background magnetic field, neglecting parallel magnetic perturbations. 
This assumption holds true when both $\beta_0 \ll 1$ and $\delta \gg 1$ \citep{scott2021turbulence}.
We define the perturbed parallel vector potential $A_1$ such that $\mathbf{b}_1=\nabla \times (A_1 \mathbf{b}_0)$.
$A_1$ is discriminated from the $A_\parallel$ that is evolved in Ohm's law \eqref{eqn:6-field:apar}.
$A_1$ represents a pure perturbation to the background magnetic field lines, while $A_\parallel$ is not a pure perturbation: it contains a part that overlaps with the magnetic equilibrium. 
This occurs because $A_\parallel$ is coupled to $j_\parallel$ via Ampere's law \eqref{eqn:6-field:amper}. 
In our `full-\textit{f}' model, $j_\parallel$ includes the Pfirsch-Schlüter current \citep{hirshman1978neoclassical}, which arises due to the balance between $\mathcal{C}\left(p_{e}+p_{i}\right)$ and $\nabla \cdot j_\parallel \mathbf{b}$ in the vorticity equation \eqref{eqn:6-field:vort}.
The Pfirsch-Schlüter current induces a shift in the magnetic equilibrium, which has already been taken into account by the equilibrium operator through tracing the background field lines. 
Thus, it should not be double-counted by the flutter operators \citep{scott2006edge}.
To extract the perturbation part of $A_\parallel$, one can remove its toroidal average (\citet{giacomin2022gbs}, GBS).
\cite{hager2020gyrokinetic} (XGC) pointed out that the transient toroidal average might include turbulent fluctuations, and therefore the time average was recommended on top of the toroidal average, but it was not applied due to the additional computational costs.
In our current study, we extract the perturbation as $A_1=A_\parallel-\langle A_\parallel \rangle_{\varphi,t}$, where the average $\langle \circ \rangle_{\varphi,t}$ is applied over both time $t$ and toroidal angle $\varphi$. 
We have found that the simulation results are not sensitive to the time averaging period $t_\mathrm{avg}$, as long as $t_\mathrm{avg}>2\pi R_0 q/V_A$, in which $V_A$ is the Alfv\'en speed and $q\approx4$ being the safety factor near the separatrix. 
Satisfying this criterion ensures that the time average can adequately smooth out the perturbations propagating along the magnetic field lines.
Otherwise, with smaller $t_\mathrm{avg}$ or without time averaging, the removed part $\langle A_\parallel \rangle_{\varphi,t}$ or $\langle A_\parallel \rangle_{\varphi}$ will exhibit poloidally rotating fluctuations, which are turbulent rather than neoclassical.

Since $A_1$ has been obtained as a turbulent perturbation, the scale of $A_1$ shall be $L_\perp$.
But the scale of $\mathbf{b}_0$, $L_B$, is  much larger, resulting in $\mathbf{b}_1=\beta_0 \nabla \times (A_1 \mathbf{b}_0)\approx \beta_0 \nabla A_1 \times \mathbf{b}_0$.
The flutter gradient of a turbulent quantity $f$ is then given by
\begin{equation}
    \mathbf{b}_1 \cdot \nabla f = -\frac{\beta_0\delta}{B_\mathrm{tor}} \mathbf{b}_0 \cdot\left(\nabla A_1 \times \nabla f\right)\,,
\end{equation}
and the flutter divergence by
\begin{equation}
     \nabla \cdot (f \mathbf{b}_1)=\mathbf{b}_1 \cdot \nabla f
     -f \nabla \cdot \mathbf{b}_1=\mathbf{b}_1 \cdot \nabla f-f B^{-1} \mathbf{b}_1 \cdot \nabla B\approx \mathbf{b}_1 \cdot \nabla f\,.
\end{equation}
The approximation is because of $ f^{-1}\nabla f \sim L_\perp^{-1} \gg B^{-1}\nabla B \sim L_B^{-1}$.
% \begin{equation}
%     \mathbf{b}_1 \cdot \nabla f = \underbrace{-\frac{\beta_0\delta}{B} \mathbf{b}_0 \cdot\left(\nabla A_1 \times \nabla f\right)}_{L_\perp^{-2}} 
%     +\underbrace{\beta_0 A_1\left(\nabla \times \frac{\mathbf{B}}{B^{2}}\right) \cdot \nabla f} _{L_\perp^{-1} L_\parallel^{-1}} 
%     -\underbrace{\beta_0 \frac{A_1}{B^{2}} \mathbf{b}_0 \cdot(\nabla B \times \nabla f)}_{L_\perp^{-1} L_\parallel^{-1}}
% \end{equation}
Based on the assumption of a strong toroidal field ($B_\mathrm{pol}\ll B_\mathrm{tor}$), the above formula can be approximated as a Poisson bracket $\mathbf{b}_0 \cdot (\nabla A_1 \times \nabla f )\approx[A_1, f]_{R,Z}$ within one poloidal plane, which can be discretized via the Arakawa scheme \citep{arakawa1997computational}.
The full system, incorporating the newly implemented magnetic flutter, has been verified by the method of manufactured solutions \citep{Salari_MMS_2000}.
\subsection{Simulation setup}\label{sec:setup}
In this study, we conducted simulations for the diverted ASDEX Upgrade (AUG) tokamak across the edge and SOL regions from $\rho=0.90$ to $\rho=1.05$ (without any down-scaling). 
Hereby, the normalised poloidal flux radius is defined as $\rho=\sqrt{(\Psi-\Psi_\mathrm{O})/(\Psi_\mathrm{X}-\Psi_\mathrm{O})}$, with $\Psi_\mathrm{O}$ and $\Psi_\mathrm{X}$ being the poloidal magnetic flux at the magnetic axis and the separatrix, respectively.
The simulation parameters were based on the attached L-mode discharge \#36190, for which GRILLIX has been previously validated \citep{zholobenko2021role}.
The toroidal magnetic field on axis in the favourable configuration is $-2.5$ T, with $\mathbf{B}\times\nabla B$ pointing towards the X-point. 
The plasma current is 800 kA. 
At the core boundary, the plasma density is $2.12\times 10^{19} \mathrm{m}^{-3}$ and the electron temperature is $300\mathrm{eV}$, as measured experimentally. 
The boundary values of plasma density and temperatures are enforced by an adaptive source at $\rho=0.90-0.92$. 
The choice of reference values is as follows: $B_0=2.5$ T, $R_0=1.65$ m, $n_0=10^{19}$ $\mathrm{m}^{-3}$, and $T_0=100$ eV. The dimensionless parameters of the system are $\delta=2854.2$, $\beta_0=3.227\times 10^{-5}$, $\mu=2.723\times 10^{-4}$, $\nu_{e 0}=12.30$, $\eta_{\parallel 0}=0.0017$, $\chi_{\parallel e 0}=940$, $\chi_{\parallel i 0}=35.35$, and $\eta_{i 0}=8.70$.
The timestep is $\mathrm{d}t = 5\times10^{-5}\tau_0$, with $\tau_0=23.84\,\mathrm{\mu s}$.
The time average period for computing $A_1$ is $t_\mathrm{avg}= 0.2\tau_0$.

It is important to acknowledge that there are 4 major free parameters in our study. These parameters are the core boundary value of ion temperature, $T_{i}^{\mathrm{core}}$, the neutrals density at the target, $N_{\mathrm{div}}$, and the free streaming parameters $\alpha^\mathrm{FS}_i$ for ion heat flux and $\alpha^\mathrm{FS}_e$ for electron heat flux.
To achieve a good match with experimental profiles and heating power, these free parameters can be fine-tuned. 
Previous GRILLIX simulations \citep{zholobenko2021role} used $\alpha^\mathrm{FS}_i=\alpha^\mathrm{FS}_e=1.0$ to obtain reasonable profiles. 
However, when we included the magnetic flutter effect as well as neoclassical ion viscosity in this present work, we noticed that using $\alpha^\mathrm{FS}_e=1.0$ leads to an excessively low electron temperature at the separatrix. 
To address this, we scan $\alpha^\mathrm{FS}_e$ at $0.1,0.3,1.0$.
% Note that even with the same $\alpha^\mathrm{FS}_{e}=\alpha^\mathrm{FS}_{i}$ the limiter is more severe for ions than for electrons because of the mass ratio.
$T_{i}^{\mathrm{core}}$ is scanned at $200\mathrm{eV}$ and $300\mathrm{eV}$.
$N_{\mathrm{div}}$ is scanned at $2\times 10^{17} \mathrm{m}^{-3}$ and $4\times 10^{17} \mathrm{m}^{-3}$.
Finally, the simulation with the combination of $\alpha^\mathrm{FS}_i=1.0$,  $\alpha^\mathrm{FS}_e=0.1$, $T_{i}^{\mathrm{core}}=200\mathrm{eV}$ and $N_{\mathrm{div}}=2\times 10^{17} \mathrm{m}^{-3}$ 
achieves the best match with the experiment and are therefore chosen as the reference.
The results and the discussion of our analysis will be based on the reference simulations.
Although the turbulence simulations are usually sensitive to those free parameters, the influence of including flutter turns out to be very robust and always consistent.
The simulation results with other free parameters are also given in the appendix as subsidiary.

\section{Results}\label{sec:results}
The reference simulations start from an initial state $t=0$ and run until \SI{3.5}{ms}.
The method for initialising profiles can be found in \citep{zholobenko2021electric}.
Both the two runs with and without flutter have reached saturation since \SI{2}{ms}. The time duration from \SI{2.098}{ms} to \SI{2.575}{ms} will be selected for statistical turbulence diagnostics throughout the entire paper if there is no additional specification. 
This interval comprises a total of 1600 samples, with 100 snapshots for each of the 16 poloidal planes.
% In Fig. \ref{fig:tracing_nt}, the time evolution of the zonally averaged density and temperatures at $\rho=0.995$ is displayed. This location is typically used to assess the saturation of turbulence. 

\subsection{A glance at stabilization by flutter}
\begin{figure}[!ht]
\centering
\begin{subfigure}[b]{0.4\textwidth}
    \includegraphics[width=\textwidth]{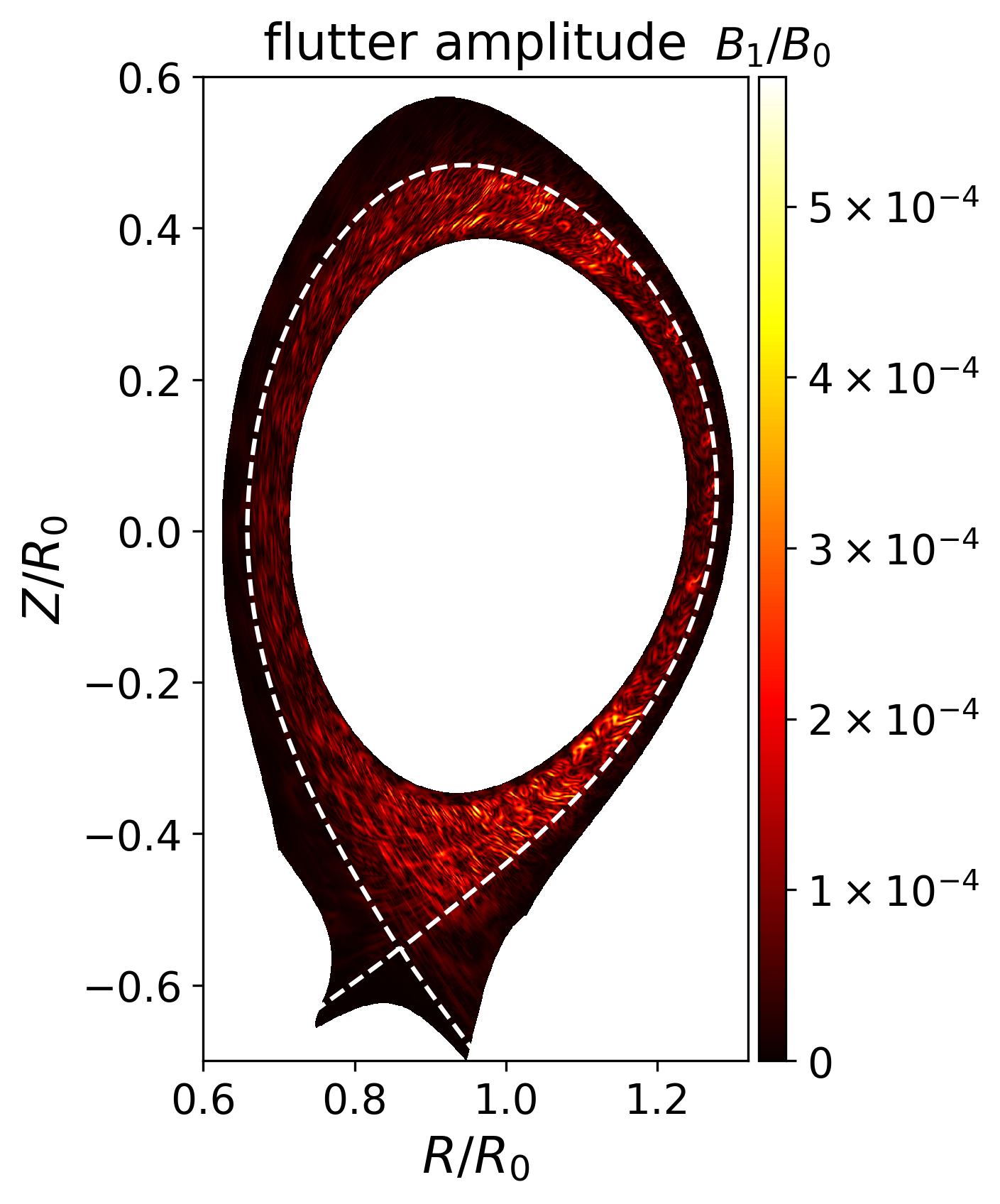}
    \caption{\label{fig:2d_b1}}
\end{subfigure}
\begin{subfigure}[b]{0.4\textwidth}
    \includegraphics[width=\textwidth]{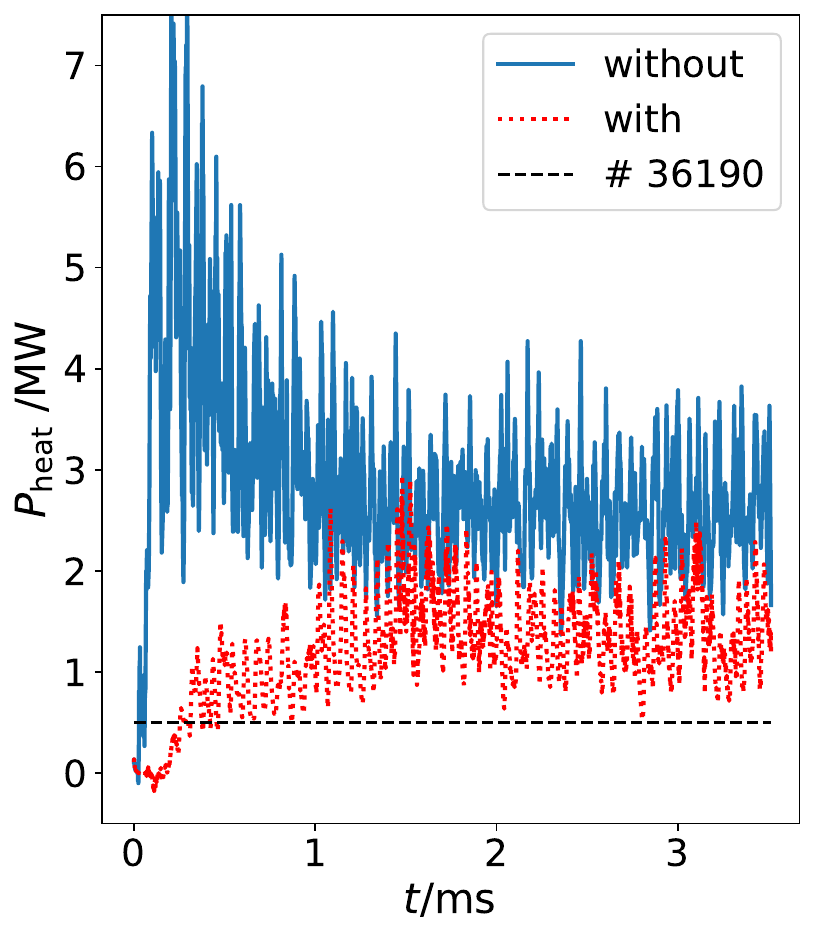}
    \caption{\label{fig:tracing_P}}
\end{subfigure}
\caption{\label{fig:stab} (a) 2D snapshot in the poloidal cross section for the instantaneous magnetic fluctuations normalised to the background field $B_1/B_0$ at $t=2.38 \mathrm{ms}$. 
The white dashed line denotes the separatrix. 
(b) Time evolution of the total injected heating power $P_\mathrm{heat}$ signaling the turbulent transport level, comparing the model without flutter (solid) and with flutter (dotted). The black dashed line is \SI{0.5}{MW} from the experiment.}
\end{figure}

Let us first get an impression of the magnetic flutter effect.
Fig. \ref{fig:2d_b1} shows the 2D snapshot depicting the normalized magnetic flutter amplitude $B_1/B_0$ at $t=2.38 \mathrm{ms}$.
It is clear that flutter activities are mainly in the confined region.
Indeed, the flutter amplitude $B_1/B_0 \sim 10^{-4}$ is minuscule, and this order of magnitude agrees with the experimental measurements \citep{mantica1991broadband}.
However, the turbulence is impressively stabilized when we include this flutter effect in the simulation. 
Fig. \ref{fig:tracing_P} illustrates the time evolution of the total heating power $P_\mathrm{heat}$ injected into the plasma. Balancing the radial heat outflow, $P_\mathrm{heat}$ serves as a reliable indicator of turbulence transport levels.
$P_\mathrm{heat}$ experiences a rapid increase before declining to saturation in the absence of flutter.
In contrast, when flutter is considered, $P_\mathrm{heat}$ shows gradual growth until saturation.
After around \SI{2}{ms}, $P_\mathrm{heat}$ saturates at approximately \SI{2.6}{MW} without flutter and \SI{1.3}{MW} with flutter. The flutter effect reduces turbulent transport by approximately half, leading to a more reasonable agreement with the experimentally observed heating power of \SI{0.5}{MW}.

\begin{figure}[!hp]
\centering
\begin{subfigure}[b]{0.75\textwidth}
    \includegraphics[width=\textwidth]{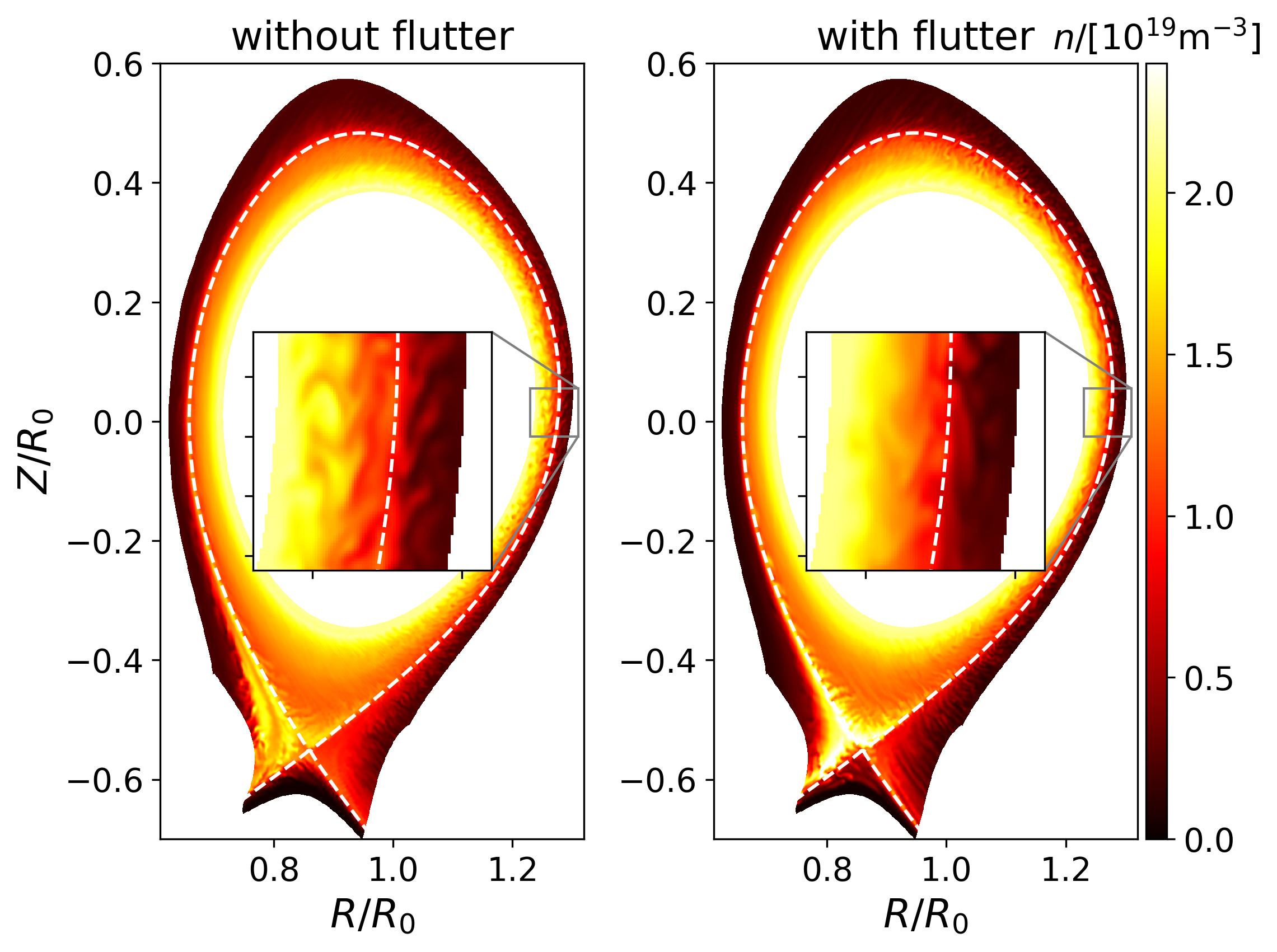}
    \caption{\label{fig:2D_ne}}
\end{subfigure}

\begin{subfigure}[b]{0.75\textwidth}
    \includegraphics[width=\textwidth]{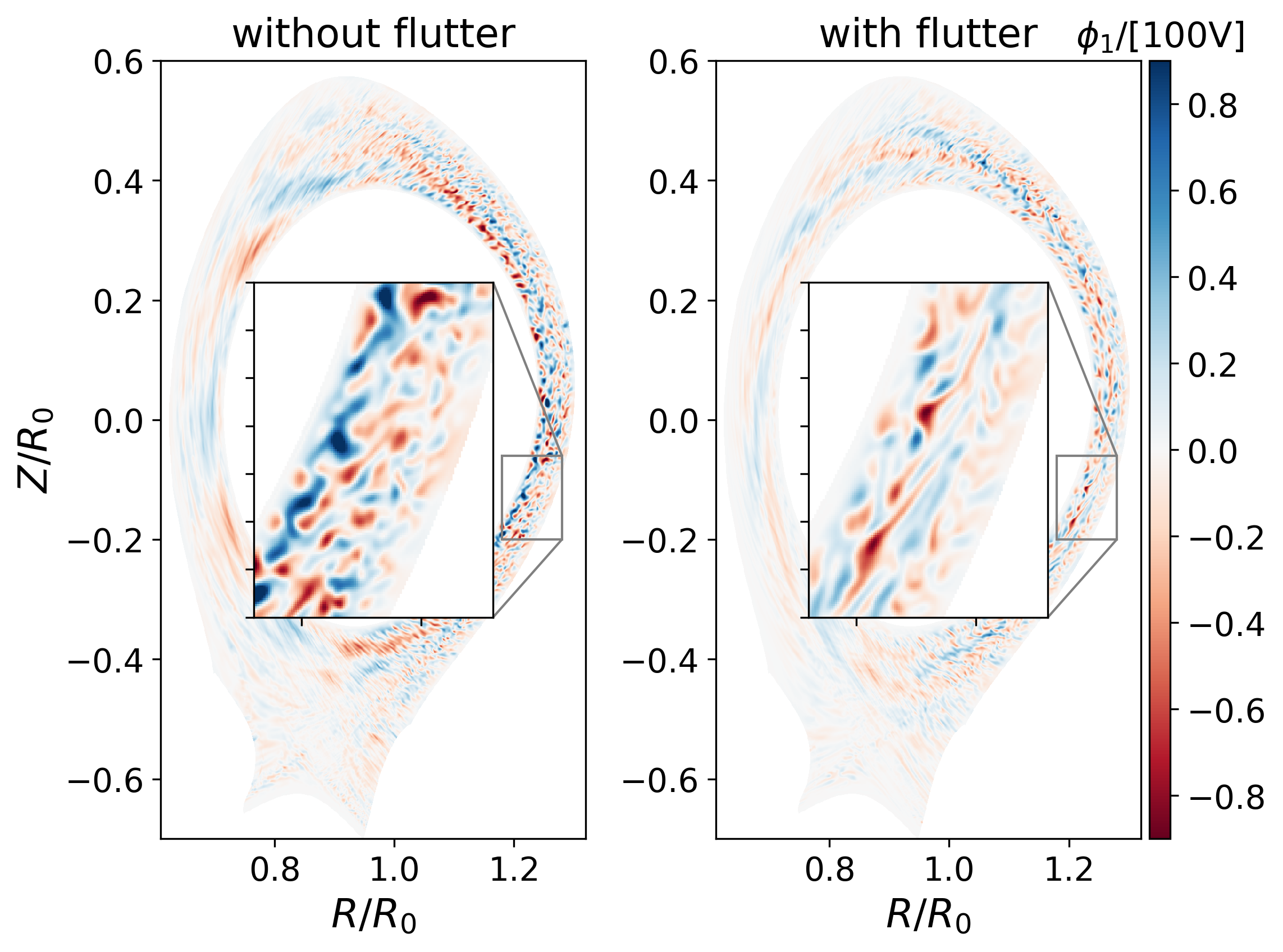}
    \caption{\label{fig:2D_pot}}
\end{subfigure}
\caption{\label{fig:2D} Snapshots for instantaneous (a) density $n$ and (b) potential fluctuation $\phi_1=\phi-\langle\phi\rangle_\varphi$, taken at $t=2.38 \mathrm{ms}$, comparing the model without flutter (left) and with flutter (right). 
The white dashed lines in (a) denote the separatrix. 
$n$ with flutter shows weaker turbulence in the edge and fewer filaments propagating into the SOL.
$\phi_1$ with flutter exhibits lower intensity and larger eddies.}
\end{figure}

To visualize the difference in the strength of turbulence, Fig. \ref{fig:2D_ne} compares the snapshots of the instantaneous density $n$ with and without flutter.
With flutter, the density is overall smoother, exhibiting gentler fluctuations and fewer turbulent filaments spreading across the separatrix.
Fig. \ref{fig:2D_pot} shows the snapshots of the instantaneous fluctuations of the potential $\phi_1=\phi-\langle\phi\rangle_\varphi$.
When flutter is included, $\phi_1$ exhibits larger eddies on the low field side. 
The poloidal correlation length \citep{zweben2002edge} gets larger with flutter, leading to the weaker Reynolds stress \citep{naulin2005shear}.
The stabilizing effect of flutter is evident.

\subsection{Validation of outboard mid-plane profiles}\label{omp}
\begin{figure}[!ht]
\centering
    \begin{subfigure}[b]{0.328\textwidth}
    \includegraphics[width=\textwidth]{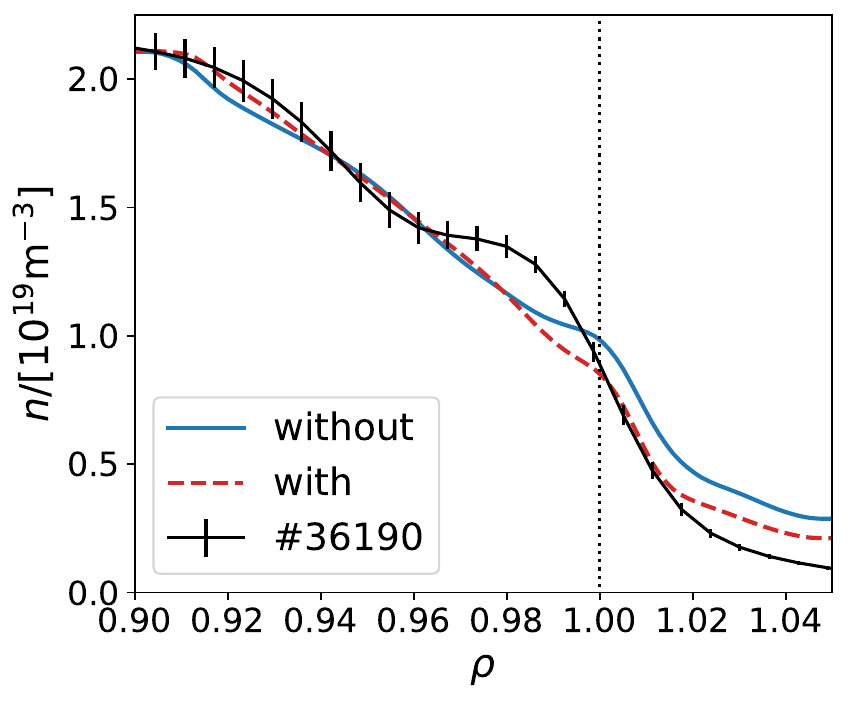}
    \caption{\label{fig:omp_avg_n}}
    \end{subfigure}
    \begin{subfigure}[b]{0.328\textwidth}
    \includegraphics[width=\textwidth]{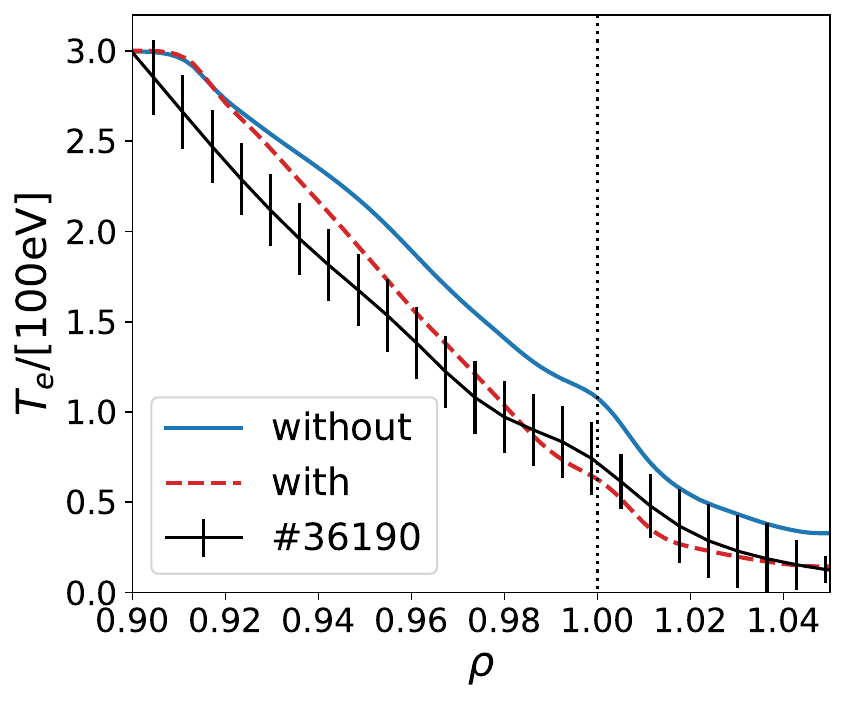}
    \caption{\label{fig:omp_avg_te}}
    \end{subfigure}
    \begin{subfigure}[b]{0.328\textwidth}
    \includegraphics[width=\textwidth]{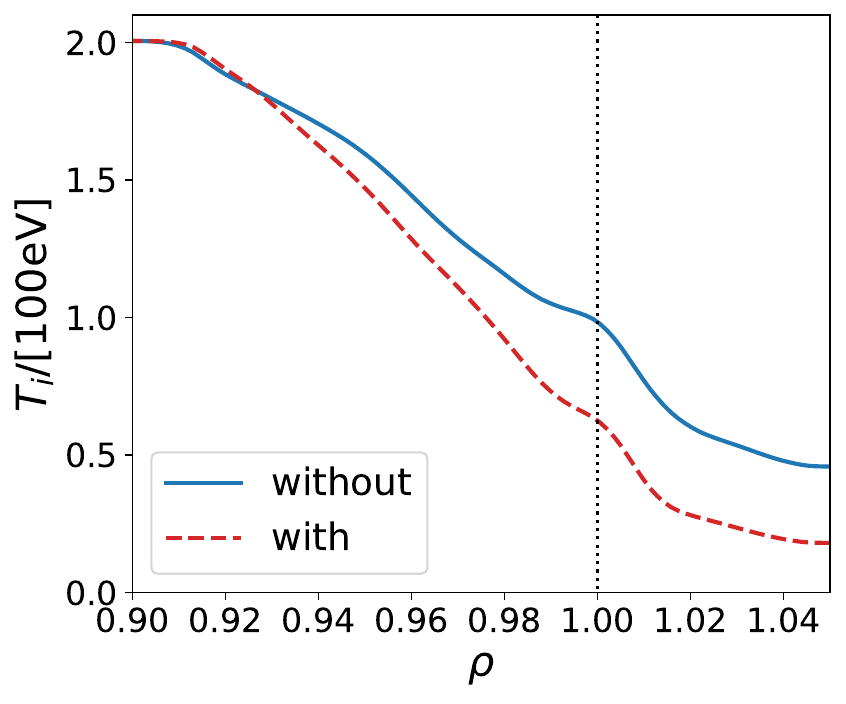}
    \caption{\label{fig:omp_avg_ti}}
    \end{subfigure}
\caption{\label{fig:omp_avg_nt} OMP profiles of time and toroidally averaged (a) plasma density, (b) electron temperature and (c) ion temperature, comparing the model without flutter (solid) and with flutter (dashed). The experiment data is from the ASDEX Upgrade discharge \#36190.}    
\end{figure}

Next, we compare our simulations against various experimental measurements from discharge \#36190. We note that results without flutter differ from our previous publication on this discharge \citep{zholobenko2021role} because also the neoclassical correction of ion viscosity has been included here, which will be discussed in more detail in a separate publication. 
In Fig. \ref{fig:omp_avg_n} and Fig. \ref{fig:omp_avg_te}, we compare the outboard mid-plane (OMP) profiles of averaged electron density $n$ and electron temperature $T_e$ with the experimental data.
Notably, the profiles of $n$ and $T_e$ demonstrate a closer match to the experimental results when flutter is taken into account, particularly in the vicinity of the separatrix. Magnetic flutter leads to steeper gradients in the confined region.
Although experimental measurements for the ion temperature $T_i$ are unavailable, the impact of flutter on $T_i$ exhibits similar characteristics as shown by Fig. \ref{fig:omp_avg_ti}. The $T_i$ profile becomes steeper in the edge and attains a lower value in the SOL when flutter is included.
It is worth noting that steeper gradients in the profiles typically indicate smaller radial particle and heat fluxes, aligning with the decreased turbulent transport levels attributed to flutter. 

\begin{figure}[!ht]
\centering
    \begin{subfigure}[b]{0.4\textwidth}
    \includegraphics[width=\textwidth]{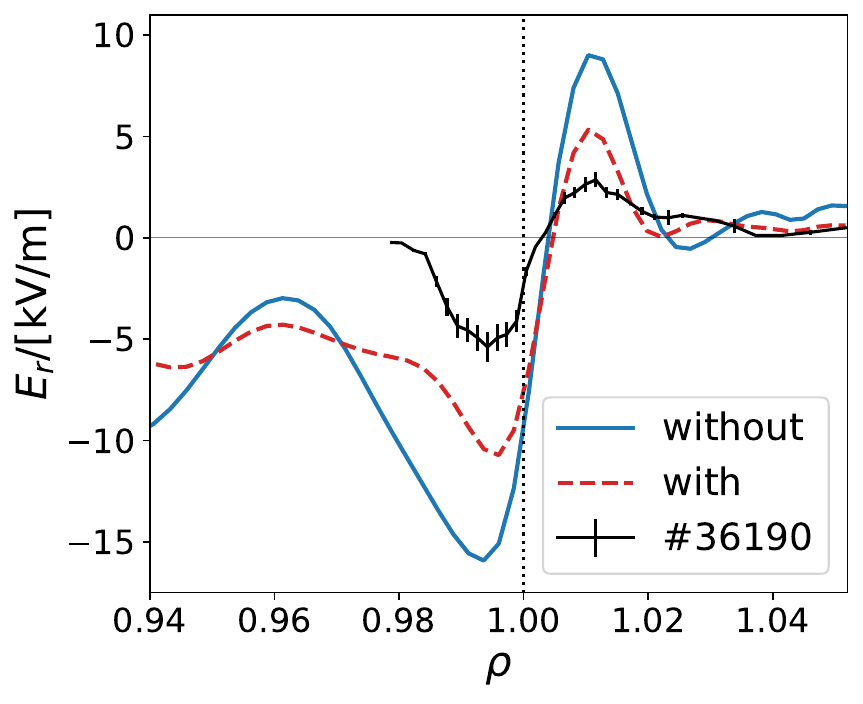}
    \caption{\label{fig:omp_avg_Er_total}}
    \end{subfigure}
    \begin{subfigure}[b]{0.4\textwidth}
    \includegraphics[width=\textwidth]{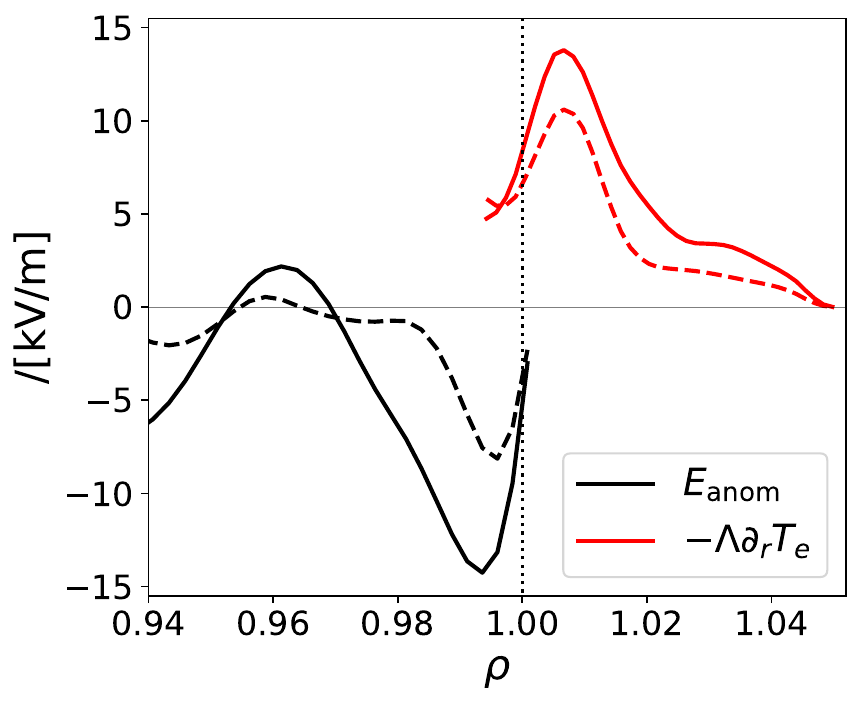}
    \caption{\label{fig:omp_avg_Er_parts}}
    \end{subfigure}
\caption{\label{fig:omp_avg_Er} OMP profiles of time and toroidally averaged (a) radial electric field $E_r=-\partial_r \phi$ and (b) radial anomalous electric fields $E_\mathrm{anom}=E_r-(n^{-1}\partial_r p_i+u_\parallel B_\mathrm{pol})$ on the edge and $-\Lambda \partial_r T_e$ on the SOL, comparing the model without flutter (solid) and with flutter (dashed). The experiment data is from the ASDEX Upgrade discharge \#36190.}    
\end{figure}

The electric field $E_r$ also shows a better agreement with the experimental data when including flutter. 
The OMP profile of $E_r$ is depicted in Fig.~\ref{fig:omp_avg_Er_total}. 
This improvement is reflected by a lower maximum value, $E_{r,\mathrm{max}}\approx 5\mathrm{kV}/\mathrm{m}$, in the SOL and a higher minimum value, $E_{r,\mathrm{min}}\approx -10\mathrm{kV}/\mathrm{m}$, in the edge.
The balance mechanism of $E_r$ varies from the SOL to edge. 
In the SOL, $E_r$ is mostly governed by $E_r\approx \Lambda \partial_r T_e$ due to sheath boundary conditions, with $\Lambda=2.69$ \citep{zholobenko2021electric}. 
Fig. \ref{fig:omp_avg_Er_parts} illustrates the variation of $\Lambda \partial_r T_e$ in the SOL, to the right side of $\rho=1$. 
With flutter, $\Lambda \partial_r T_e$ decreases, leading to a lower $E_{r,\mathrm{max}}$.
Therefore, the better agreement of $E_{r}$ in the SOL can be attributed to the improved match of $T_e$ profile as depicted in Fig.~\ref{fig:omp_avg_te}.
In the edge, $E_r=n^{-1}\partial_r p_i+u_\parallel B_\mathrm{pol}+E_\mathrm{anom}$ is balanced by the ion pressure gradient force $n^{-1}\partial_r p_i$, the Lorentz force $u_\parallel B_\mathrm{pol}$, and the anomalous electric field $E_\mathrm{anom}$.
The first two parts are described by the neoclassical theory \citep{helander2005collisional}, while $E_\mathrm{anom}$ is introduced by the Reynolds stress in turbulence simulations \citep{zholobenko2021electric}.
When flutter is included, the radial ion pressure gradient becomes steeper as indicated by Fig. \ref{fig:omp_avg_nt}, which does not explain the higher $E_{r,\mathrm{min}}$.
Rather, the improved match of $E_{r,\mathrm{min}}$ is found to result from the weaker anomalous electric field $E_\mathrm{anom}$, which is shown by the black lines in Fig.~ \ref{fig:omp_avg_Er_parts}.
A weaker $E_\mathrm{anom}$ implies a smaller Reynolds stress and weaker turbulence, as previously observed in Fig. \ref{fig:2D_pot}.
% The correlation length, roughly the size of the largest eddies in the turbulent flow, is larger in the case with flutter, leading to the weaker Reynolds stress and the weaker turbulence.

% In conclusion, including flutter lowers the heating power and consistently improves the OMP validation results.
% Our code becomes more capable of prediction with the incorporation of flutter.
% \begin{figure}[!htbp]
% \centering
%     \begin{subfigure}[b]{0.8\textwidth}
%     \includegraphics[width=\textwidth]{figs/011_omp_std.pdf}
%     \caption{\label{fig:omp_std}}
%     \end{subfigure}

%     \begin{subfigure}[b]{0.8\textwidth}
%     \includegraphics[width=\textwidth]{figs/011_omp_skew.pdf}
%     \caption{\label{fig:omp_skew}}
%     \end{subfigure}
    
%     \begin{subfigure}[b]{0.8\textwidth}
%     \includegraphics[width=\textwidth]{figs/011_omp_kurt.pdf}
%     \caption{\label{fig:omp_kurt}}
%     \end{subfigure}    
% \caption{\label{fig:omp_fluct} OMP profile of (a) normalised standard deviation, (b) skewness and (c) kurtosis, comparing the model without flutter (left) and with flutter (right)}
% \end{figure}

\subsection{Radial turbulent fluxes}\label{sec:zonal}
The OMP profiles evolve according to the self-consistent turbulent fluxes, which might be changed by flutter and will be examined in this section.
In our system, the radial particle flux $\Gamma$ for electrons going across magnetic flux surfaces reads
\begin{equation}
    \label{eqn:Gamma}
    \underbrace{\Gamma}_{\mathrm{total}}=
    \underbrace{n\mathbf{v}_E\cdot\mathbf{e}_\rho}_{\mathrm{ExB}}
    + \underbrace{n v_\parallel \mathbf{b}_1\cdot\mathbf{e}_\rho}_{\mathrm{mag}} 
    +\underbrace{(-\frac{\mathbf{b}_0 }{B_\mathrm{tor}} \times \nabla p_e)\cdot\mathbf{e}_\rho}_{\mathrm{dia}}\,,
\end{equation}
where $v_\parallel=u_\parallel-j_\parallel/n$ is the electron parallel velocity, and $\mathbf{e}_\rho$ is the radial unit vector.
Without flutter, the flux $\Gamma$ is driven by the $E \times B$ and diamagnetic drifts. However, in the presence of flutter the parallel flux of electrons $nv_\parallel$ can also be projected perpendicularly due to the perturbation $\mathbf{b}_1$. This additional term is referred to as the magnetic flutter flux and is denoted as `mag'.
The radial electron and ion heat fluxes are
\begin{equation}
    \label{eqn:Qe}
    \underbrace{Q_e}_{\mathrm{total}}=
    \underbrace{\frac{5}{2}nT_e \mathbf{v}_E\cdot\mathbf{e}_\rho}_{\mathrm{ExB}} 
    + \underbrace{(\frac{5}{2}nT_e v_\parallel 
    + q_{e,\parallel}  - 0.71 j_\parallel T_e )\mathbf{b}_1\cdot\mathbf{e}_\rho
    }_{\mathrm{mag}}
    + \underbrace{(-\frac{5\mathbf{b}_0 }{2 B_\mathrm{tor}} \times \nabla p_e T_e)\cdot\mathbf{e}_\rho}_{\mathrm{dia}}\,,
\end{equation}
\begin{equation}
    \label{eqn:Qi}
    \underbrace{Q_i}_{\mathrm{total}}=
    \underbrace{\frac{5}{2}nT_i \mathbf{v}_E\cdot\mathbf{e}_\rho}_{\mathrm{ExB}} 
    + 
    \underbrace{(\frac{5}{2}nT_i u_\parallel
    + q_{i,\parallel})\mathbf{b}_1\cdot\mathbf{e}_\rho
    }_{\mathrm{mag}}
    % + \frac{5}{2}nT_i \mathbf{u}_\mathrm{dia}\cdot\mathbf{e}_\rho
    % +\frac{5}{2}nT_i\mathbf{u}_\mathrm{pol}\cdot\mathbf{e}_\rho
    +\underbrace{(\frac{5 \mathbf{b}_0 }{2 B_\mathrm{tor}}\times \nabla p_i T_i)\cdot\mathbf{e}_\rho}_{\mathrm{dia}}\,,
\end{equation}
% Note that the heat flux driven by the ion polarization drift does not appear.
With the presence of magnetic flutter, the parallel heat advection terms $\frac{5}{2}nT_e v_\parallel$ and $\frac{5}{2}nT_i u_\parallel$, as well as the parallel heat conduction terms $q_{e, \parallel}$ and $q_{i, \parallel}$ (and an additional term $- 0.71 j_\parallel T_e$ for electrons due to thermal force), can traverse the magnetic flux surfaces. Collectively, these terms are denoted by `mag'.

\begin{figure}[!htbp]
\centering
    \begin{subfigure}[b]{0.328\textwidth}
    \includegraphics[width=\textwidth]{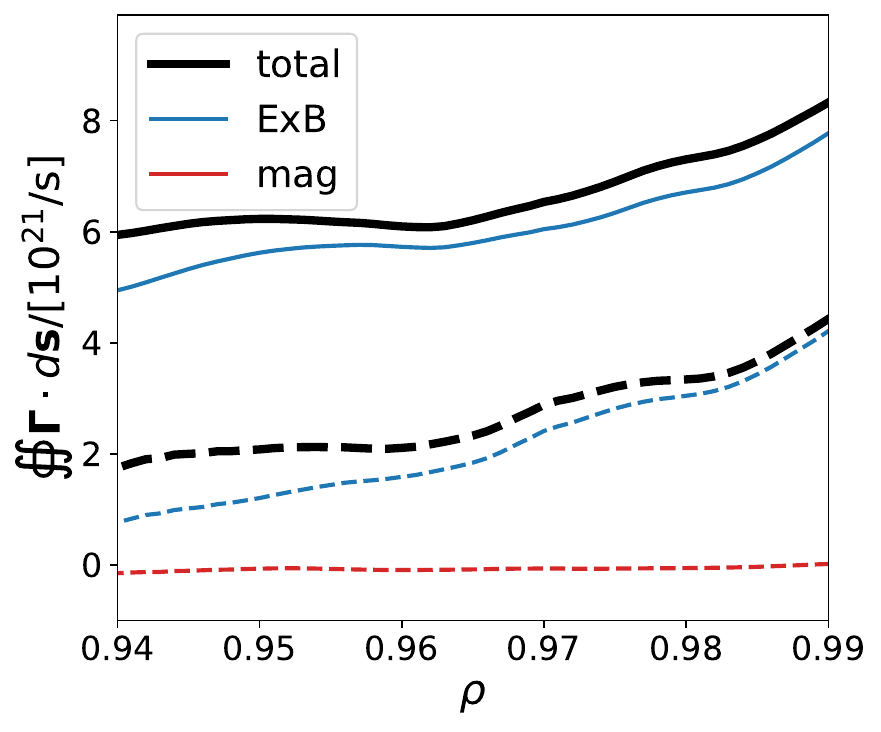}
    \caption{\label{fig:zonal_flux_n}}
    \end{subfigure}
    \begin{subfigure}[b]{0.328\textwidth}
    \includegraphics[width=\textwidth]{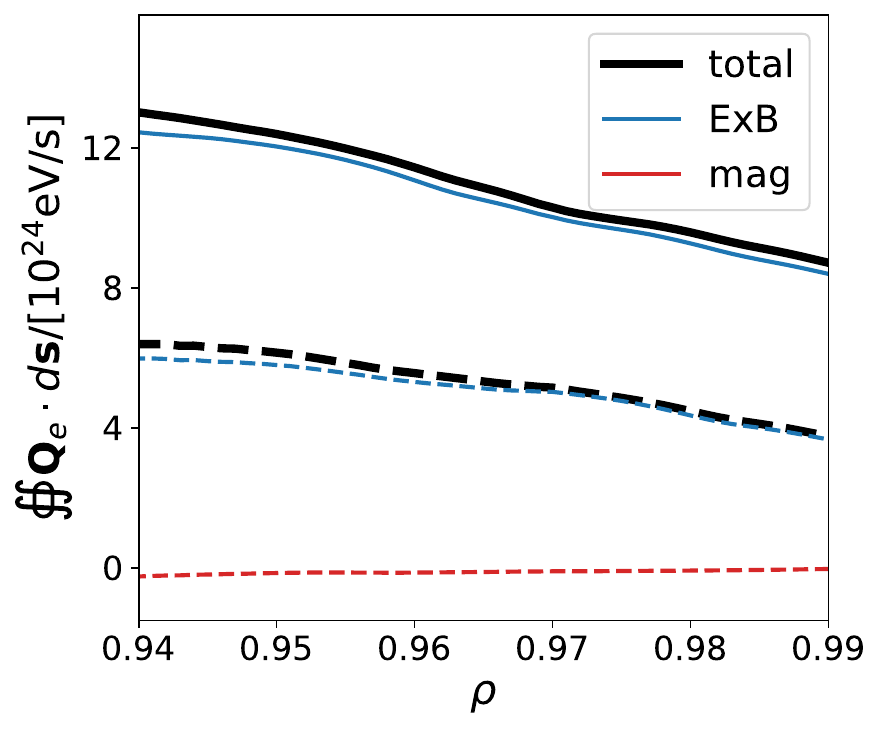}
    \caption{\label{fig:zonal_flux_Te}}
    \end{subfigure}
    \begin{subfigure}[b]{0.328\textwidth}
    \includegraphics[width=\textwidth]{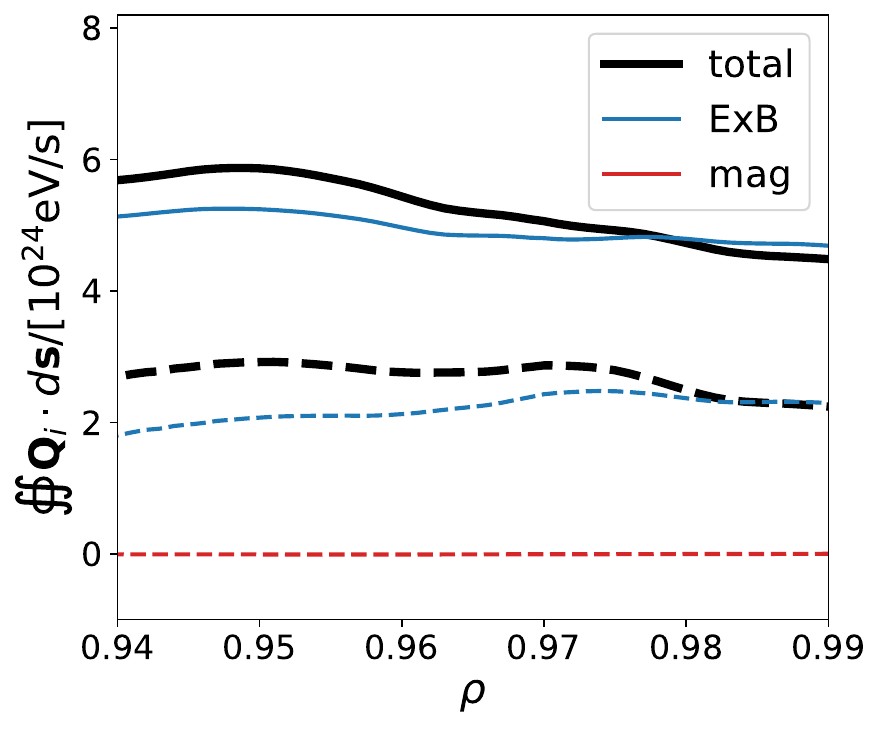}
    \caption{\label{fig:zonal_flux_Ti}}
    \end{subfigure}
\caption{\label{fig:zonal_flux} Surface integrals of (a) density flux, (b) electron heat flux, and (c) ion heat flux crossing the magnetic flux surfaces as a function of $\rho$, signed positive for outward fluxes. 
The total flux `total', $E \times B$ flux `ExB' and magnetic-flutter flux `mag' are defined by Eqn.\eqref{eqn:Gamma}-\eqref{eqn:Qi} and compared for the models without flutter (solid) and with flutter (dashed).
Note that the model without flutter has no `mag' flux.}  
\end{figure}

Fig. \ref{fig:zonal_flux} compares these radial fluxes in terms of their surface integrals of the components perpendicular to the magnetic flux surfaces.
It is evident that the total particle flux and electron and ion heat fluxes all significantly decrease with flutter, as indicated by the black lines. 
The reduction in fluxes implies weaker turbulent transports, which explains the lower heating power in Fig. \ref{fig:tracing_P} and aligns with changed OMP profiles in Fig. \ref{fig:omp_avg_nt}.

The influence of flutter on the fluxes is indirect. 
The particle and heat fluxes directly caused by magnetic flutter (referred to as `mag' and shown by the red lines) are roughly two orders of magnitude below the total fluxes, conforming to the experimental measurements \citep{mantica1991broadband} and the previous electromagnetic simulations \citep{scott1997three,naulin2003electromagnetic,mandell2020electromagnetic,giacomin2022turbulent}.
% This is primarily because the inductivity $\partial A_\parallel / \partial t$ is too small to build up its own phase shift between $A_\parallel$ and $v_\parallel$ or $q_\parallel$ at low $\beta$ \citep{scott2021turbulence}. 
For the electron heat flux, the flutter contribution is even slightly negative, which agrees with previous local simulations at low $\beta$ \citep{jenko1999numerical,scott2021turbulence2}.
The insignificance of flutter fluxes also suggests that neither the kinetic ballooning mode (KBM) \citep{pueschel2010transport} nor the microtearing mode (MTM) \citep{doerk2011gyrokinetic} is triggered on any flux surfaces in this L-mode simulation.
Nevertheless, the indirect effect of flutter on the $E \times B$ fluxes (blue lines) is notable: the $E \times B$ fluxes are reduced dramatically, which is the primary reason for the overall reduction in total fluxes. 
This flutter effect of decreasing $E \times B$ fluxes has already been predicted by the local model for collisionless Drift Alfv\'en turbulence \citep{jenko1999numerical} in H-mode AUG conditions.
However, quantitatively, it is not obvious that this flutter effect is so strong that up to 50\% of the $E \times B$ fluxes can be canceled even in a L-mode AUG simulation, especially both for the electrons and for the ions. 
This highlights the significance of magnetic flutter in shaping the turbulent transport behavior. 

To further investigate the behaviour of heat fluxes, we split the averaged $E \times B$ heat fluxes into an advective and a conductive component as shown in Eqn. \eqref{eqn:ExB_2parts},
\begin{equation}\label{eqn:ExB_2parts}
    \langle\frac{5}{2}nT_{e,i} \tilde{v}_E\rangle=
    \underbrace{\frac{5}{2}\langle T_{e,i}\rangle \langle \tilde{n} \tilde{v}_E \rangle}_{\mathrm{adv}}
    + 
    \underbrace{\frac{5}{2}\langle n \rangle \langle  \tilde{T}_{e,i} \tilde{v}_E \rangle}_{\mathrm{cond}}\,,
\end{equation}
where $\mathbf{v}_E\cdot\mathbf{e}_\rho=\tilde{v}_E$, and $\langle \tilde{v}_E \rangle_{t,\varphi}=0$ is assumed. Triple correlations are also neglected. 
The advective part is primarily determined by the correlation between density fluctuations and $E \times B$ fluctuations, which corresponds to the $E \times B$ particle flux. In the tokamak core, the advective part is usually small due to a small particle flux, and the heat flux is mainly conductive. However, in the edge region, the advective and conductive parts can be comparable.

\begin{figure}[!htbp]
\centering
\begin{subfigure}[b]{0.4\textwidth}
    \includegraphics[width=\textwidth]{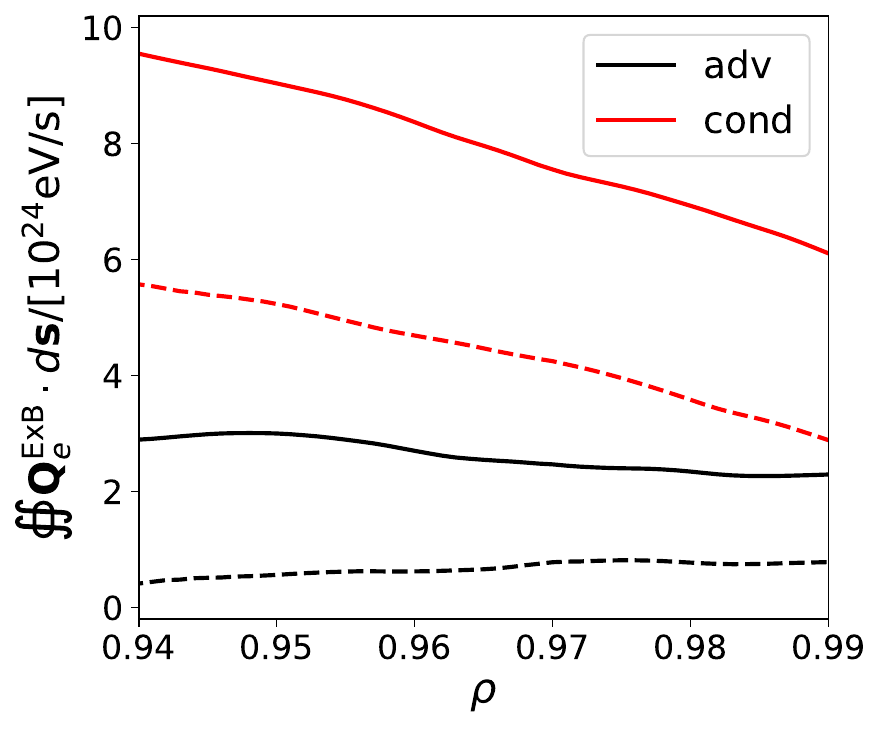}
    \caption{\label{fig:zonal_flux_ExB_Te}}
\end{subfigure}
\begin{subfigure}[b]{0.4\textwidth}
    \includegraphics[width=\textwidth]{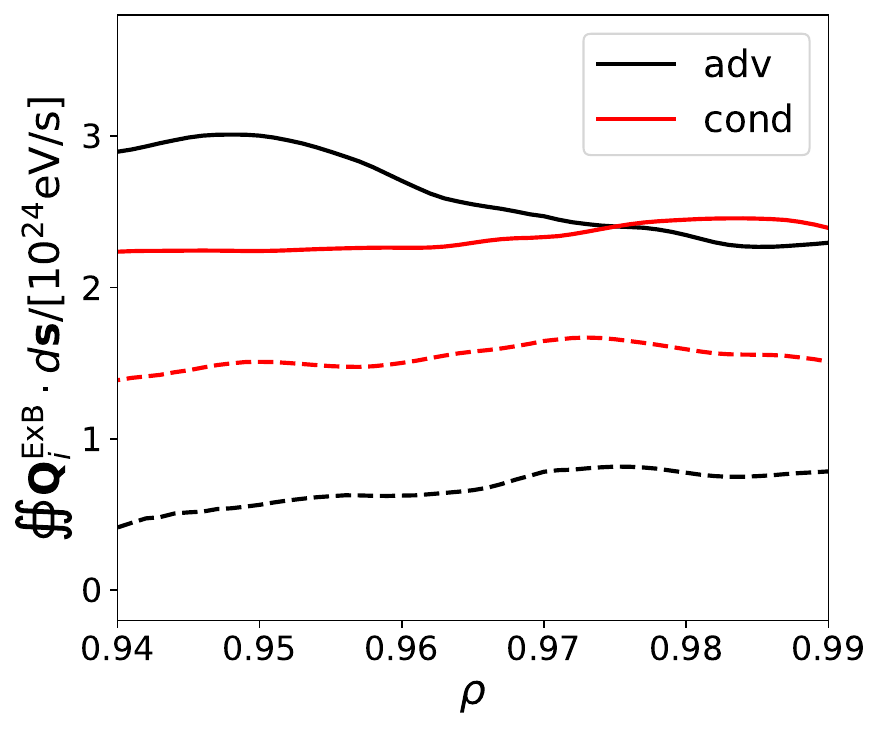}
    \caption{\label{fig:zonal_flux_ExB_Ti}}
\end{subfigure}
\caption{\label{fig:zonal_flux_ExB_T} Surface integrals of the advective (`adv') and conductive (`cond') components of (a) electron $E \times B$ heat flux and (b) ion $E \times B$ heat flux, calculated by Eqn.\eqref{eqn:ExB_2parts}, comparing the model without flutter (solid lines) and with flutter (dashed lines). Both the `adv' and `cond' are reduced when flutter is present.}
\end{figure}

The effect of magnetic flutter on $E \times B$ heat fluxes is observed to be twofold, decreasing both the advective and conductive components. 
Fig.~\ref{fig:zonal_flux_ExB_T} illustrates the changes in the advective and conductive $E \times B$ heat fluxes made by flutter. 
It is observed that:
(1) both the $E \times B$ heat advection $Q^\mathrm{adv}_{e,i}$ and conduction $Q^\mathrm{cond}_{e,i}$ are decreased by flutter.
This implies that the magnetic flutter not only intervenes in the correlation between $E \times B$ and density fluctuations, but also between $E \times B$ and temperature fluctuations.
(2) The conduction decreases more significantly than advection does for the electron $E \times B$ heat flux, but quite the opposite for ions.
The ion $E \times B$ heat conduction $Q^\mathrm{cond}_i$ and advection $Q^\mathrm{adv}_e$ have a similar magnitude in the case without flutter.
When flutter is included, $Q^\mathrm{cond}_i$ becomes about twice $Q^\mathrm{adv}_i$, coming as the secondary result on top of the overall reduced $E \times B$ heat flux.
% We notice three interesting points in Fig.~\ref{fig:zonal_flux_ExB} as followings
% (1) With flutter, both the advective (green lines) and conductive (blue lines) parts are reduced.
% (2) For both with/without flutter cases, the advective parts are always smaller than conductive part in the whole confined region, indicated by the red lines with values smaller $1$. But the advective parts gain significance gradually close to the separatrix, indicating the growing drift wave instability near the open field line region.
% (3) The reduction by flutter in advective ExB heat fluxes is more dramatic than the conductive counterparts.
% As a result, the proportions of advective to conductive parts also decrease with flutter.
% In other words, the ExB heat fluxes become more conductive with flutter.
% This agrees with our previous conclusion that with flutter the turbulent transports become more likely driven by ITG instability as indicated by Figs.\ref{fig:omp_etai} and \ref{fig:omp_std}.

\subsection{Fourier spectra analysis}
\begin{figure}[!htbp]
\centering
    \includegraphics[width=0.5\textwidth]{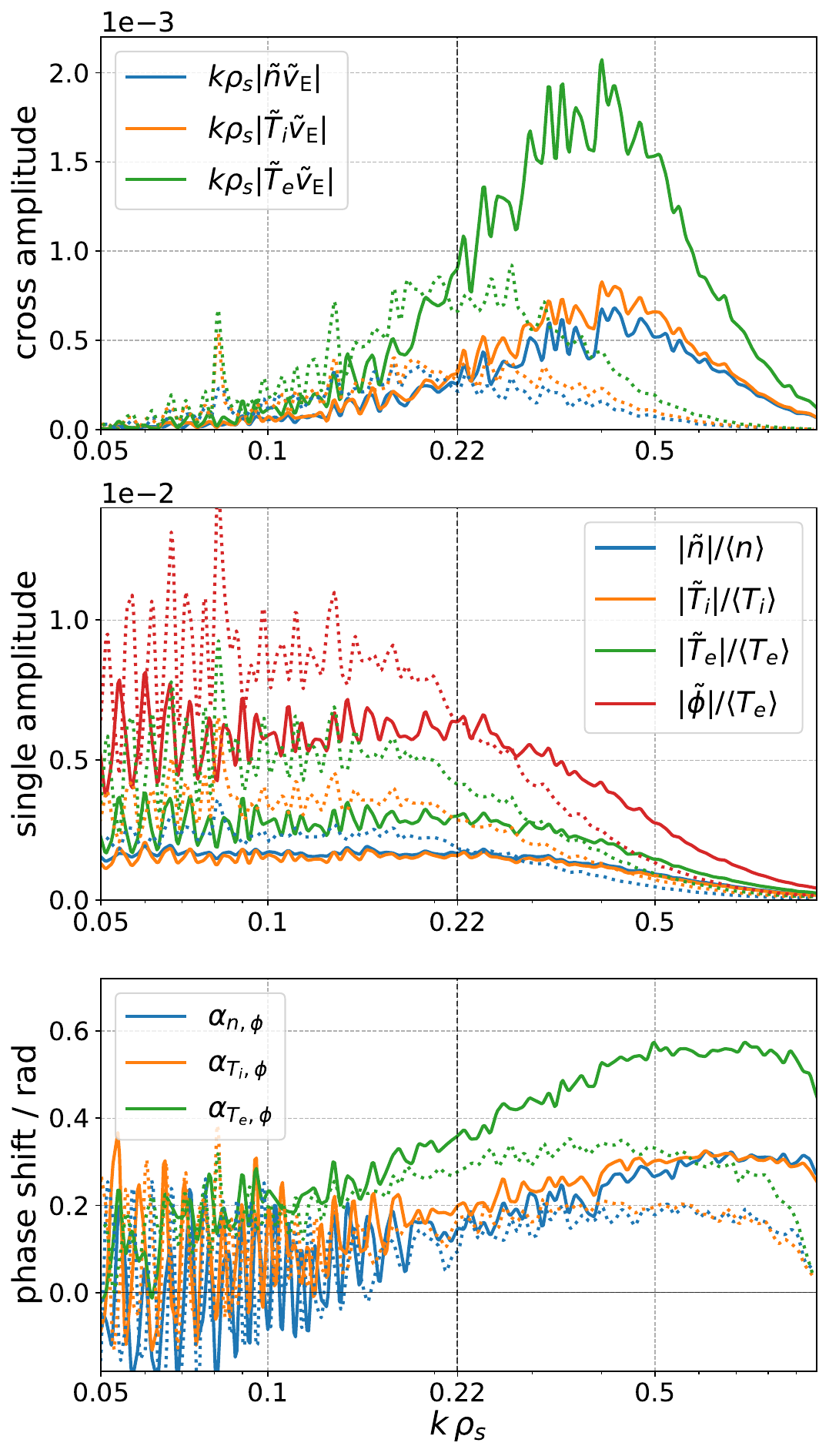}
\caption{\label{fig:fourier_all_rho990} Fourier spectra at $\rho=0.98$ for (a) the cross fluxes $| \tilde{n} \tilde{v}_E |$, $|\tilde{T}_{e} \tilde{v}_E|$ and $|\tilde{T}_{i} \tilde{v}_E|$ multiplied by $k\rho_s$ (b) the fluctuation amplitudes of individual quantities $|\tilde{\phi}|$, $|\tilde{n}|$ and $|\tilde{T}_{e,i}|$, normalised to the time and zonally averaged values, and (c) the phase shifts (in radians) with respect to potential fluctuation,
comparing the model without flutter (solid) and with flutter (dotted).}
\end{figure}

In the previous chapter, we observed that radial $E \times B$ fluxes were reduced by magnetic flutter in terms of their zonal integrals. However, it is also interesting to examine how flutter alters the spectral distribution of fluxes in the Fourier space.
For this purpose, we performed Fourier analysis for the cross spectra $|\tilde{n} \tilde{v}_E|$ (responsible for the particle flux and advective heat fluxes), $|\tilde{T}_{e} \tilde{v}_E|$ and $|\tilde{T}_{i} \tilde{v}_E|$ (responsible for electron and ion conductive heat fluxes) on a closed magnetic flux surface with $\rho=0.98$. The results are shown in Fig.~\ref{fig:fourier_all_rho990} (top).
The x-axis of the plot represents the poloidal wave number $k$ multiplied by the normalized Larmor radius $\rho_s=\rho_{s0}\sqrt{\langle T_e \rangle}$, which depends on the time and zonally averaged electron temperature $\langle T_e \rangle$ at $\rho=0.98$. Note that a logarithmic x-scale is used, which requires the flux quantities in this plot to be multiplied by a factor of $k\rho_s$ to accurately represent the transport (the area under the curve is the total flux).
From the plot, we can see that the main spectral range contributing to the fluxes slightly shifts towards larger scales from $k\rho_s\sim0.4$ without flutter to $k\rho_s\sim0.2$ with flutter. 
All the fluxes are dramatically decreased by flutter on small scales but slightly increased on large scales. 
Therefore, we conclude that flutter reduces $E \times B$ fluxes mainly due to the reduction in small scales.

The influence of flutter on the $E \times B$ fluxes is a consequence of the altered turbulent dynamics of fluctuating density, temperatures, and potential, and their correlations among each other. 
These radial $E \times B$ fluxes for a given wave number $k$ in Fourier spaces can be decomposed as follows \citep{zholobenko2023filamentary}:
\begin{equation}
    \label{eqn:ExB_decomp}
    \Gamma\sim k |\tilde{n}| |\tilde{\phi}| \sin(\alpha_{n,\phi})\,,
    \quad
    Q_{e}^\mathrm{cond}\sim k |\tilde{T}_{e}| |\tilde{\phi}| \sin(\alpha_{T_e,\phi})\,,
    \quad
    Q_{i}^\mathrm{cond}\sim k |\tilde{T}_{i}| |\tilde{\phi}| \sin(\alpha_{T_i,\phi})\,.
\end{equation}
The phase shift angle $\alpha$ between any two fluctuations $\tilde{A}$ and $\tilde{B}$ in the complex domain is defined by $\alpha_{A,B}=\mathrm{Im} (\log{\tilde{A}\tilde{B}^*})$.
In Fig. \ref{fig:fourier_all_rho990} (middle), the Fourier amplitudes of individual quantities ($|\tilde{n}|$ and $|\tilde{T}_{e,i}|$ normalized to their time and zonal average $\langle \circ \rangle$, and $|\tilde{\phi}|$ normalized to $\langle T_e \rangle$) are plotted on the same x-axis. Similar to the fluxes, flutter decreases the amplitudes of all quantities on small scales but increases them on large scales.
The phase shifts are plotted in Fig. \ref{fig:fourier_all_rho990} (bottom).
We notice that the phase shifts with respect to the potential fluctuation $\tilde{\phi}$ are all appreciably decreased by flutter on small scales. This observation suggests that flutter expedites the response of $\tilde{\phi}$ to $\tilde{n}$, $\tilde{T}_i$, and $\tilde{T}_e$ simultaneously. 
As a result, the radial $E \times B$ heat advection and conduction are both reduced, explaining the results in Fig.~\ref{fig:zonal_flux_ExB_T}.

By examining the three plots together in Fig. \ref{fig:fourier_all_rho990}, we observe that $k\rho_s\sim0.22$ serves as a meaningful dividing line. 
On the right side of $k\rho_s\sim0.22$, all flux amplitudes, phase shifts, and $\tilde{n}$ and $\tilde{\phi}$ amplitudes are reduced synchronously, although the reductions in $\tilde{T}_{e,i}$ amplitudes are shifted to even smaller scales. 
The smaller value of $\alpha_{n,\phi}$ is a clear evidence for more adiabatic electrons and drift-wave-like turbulence.
For drift-wave turbulence, as will be explained in detail down below, the linear stabilization is associated with a reduced phase shift, which in turn leads to smaller fluctuation amplitudes. 
All these factors contribute to the reduction in cross amplitudes of transport. 
Therefore, understanding the role of flutter in the phase shift is crucial for interpreting the overall decrease in transport caused by flutter.

Overall the dynamics of temperatures and density show similar responses to flutter, but there are still notable differences.
We notice that
(1) the reduction in $\alpha_{T_e,\phi}$ is more significant than in $\alpha_{n,\phi}$. This indicates that the response of potential fluctuations to electron temperature fluctuations is more expedited by flutter compared to the response to density fluctuations. 
This explains why the reduction in $E \times B$ conduction for electrons is more pronounced than the reduction in $E \times B$ advection as shown in Fig.~\ref{fig:zonal_flux_ExB_Te}.
(2) without flutter we find $|\tilde{n}|\approx|\tilde{T}_i|$, whereas with flutter the amplitude of $|\tilde{T}_i|$ exceeds $|\tilde{n}|$ especially on large scales $k\rho_s<0.5$. 
The higher fluctuation amplitudes of $|\tilde{T}_i|$ can compensate for the reduction in $E \times B$ conduction caused by $\alpha_{T_i,\phi}$ decreasing. 
This is why the ion $E \times B$ conduction decreases slightly less than the advection does in Fig.~\ref{fig:zonal_flux_ExB_Ti}.

\section{Discussion}\label{sec:discussion}
To understand the effect of flutter on edge turbulence, we need to investigate its impact on the micro-instabilities that drive the turbulence. 
The Fourier spectral analysis has shown that the core role of flutter is to uniformly reduce all the phase shifts $\alpha_{n,\phi}$, $\alpha_{T_i,\phi}$, and $\alpha_{T_e,\phi}$, which suggests a stabilizing effect on the drift-wave (DW) instability. However, the amplified $|\tilde{T}_i|$ indicates a destabilizing effect on the ITG instability. 
In the edge region, turbulence is usually complex, with a mixture of modes or instabilities, which makes it challenging to isolate the individual effects of flutter on DW and ITG.
This chapter is thereby dedicated to identifying the role of flutter in micro-instabilities. 
In section \ref{sec:regimes}, we establish that the effect on ITG is a secondary outcome of the stabilized DW turbulence, while the effect on the DW instability is intrinsic and primary. 
In section \ref{sec:linear}, we develop a linear model for the drift-Alfv\'en-wave (DAW) instability, enabling us to explain the influence of flutter analytically. 
In section \ref{sec:numexp}, we conduct numerical experiments with varied subsets of flutter terms confirming that flutter mainly works in DAW-relevant terms.
Additionally, we discuss the nonlinear stabilization by flutter and the dependence on $\beta$.

\subsection{Changes in turbulence regimes}\label{sec:regimes}
The competition among various instability mechanisms can be reflected in the fluctuation amplitudes at the OMP. This work represents the amplitude by the standard deviation normalized against the mean value. 
For example, the density fluctuation amplitude is given by $A(n) = \sigma(n) / \langle n \rangle_{t,\varphi}$.
The treatment for potential fluctuations is special. Firstly, the potential fluctuations are normalized by the mean electron temperature, as the averaged potential could be zero somewhere. Secondly, the potential fluctuations are very high throughout due to flux-surface-symmetric potential oscillations, which are due to Geodesic Acoustic Modes (GAM) rather than ballooning modes. To exclude the GAM oscillation, we remove the toroidal average of the potential by defining $\phi_1=\phi-\langle \phi \rangle_\varphi$. (In principle, one should remove the zonal average 
of $\phi$, but this cannot be easily applied to the open field line region.)

\subsubsection{Effect on drift-wave}
Figure \ref{fig:omp_fluct_phi_pei} presents the OMP profile of fluctuation amplitudes for $\phi_1$, $p_e$, and $p_i$. 
The drift-wave turbulence is characterized by $A(\phi_1)\sim A(p_e)\sim A(p_i)$ \citep{scott2005drift,manz2018microscopic}.
With the inclusion of flutter, the amplitude $A(\phi_1)$ decreases, while $A(p_e)$ and $A(p_i)$ increase, resulting in the amplitudes becoming closer to each other. 
This observation suggests that flutter makes the turbulence more drift-wave-like. 
Therefore, investigating the role of flutter in the drift-wave instability is justified.
Despite the turbulence becoming more drift-wave-like, the transport driven by the drift-wave instability is actually lower with flutter. This reduced particle and heat transport has been demonstrated by the smaller phase shifts shown in Fig. \ref{fig:fourier_all_rho990}. 
% For linear DW instability, a smaller phase shift is associated with a lower growth rate. 
In this sense, flutter can stabilize the drift-wave instability.

\subsubsection{Effect on ITG}
\begin{figure}[!htbp]
\centering
\begin{subfigure}[b]{0.4\textwidth}
    \includegraphics[width=\textwidth]{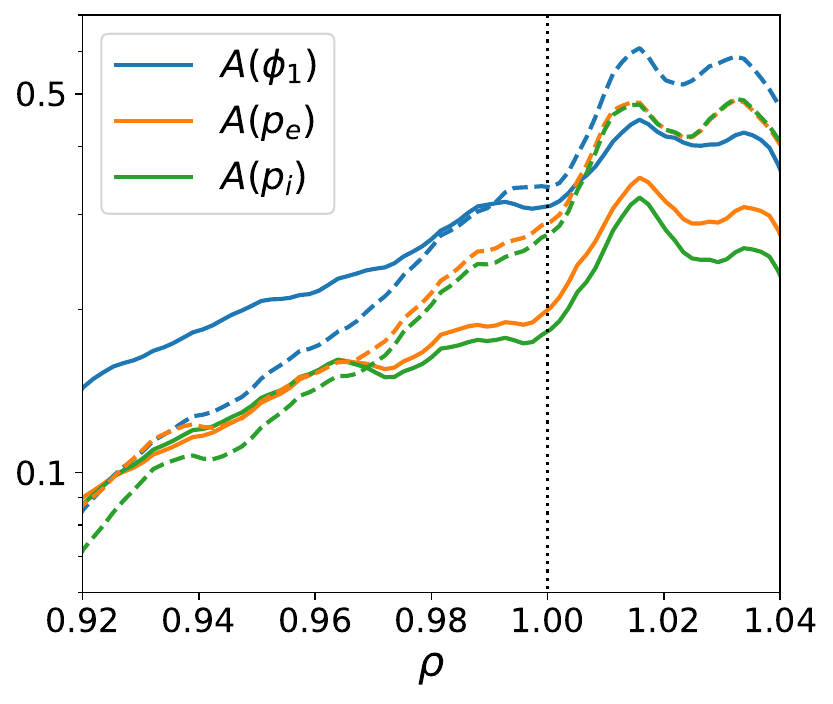}
    \caption{\label{fig:omp_fluct_phi_pei}}
\end{subfigure}
\begin{subfigure}[b]{0.4\textwidth}
    \includegraphics[width=\textwidth]{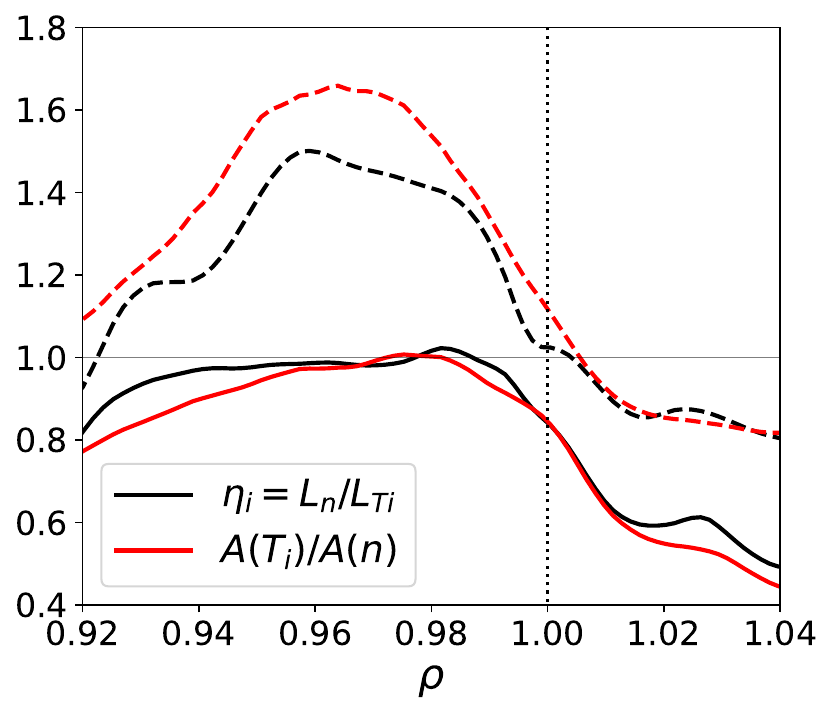}
    \caption{\label{fig:omp_eta_Ti}}
\end{subfigure}
\caption{\label{fig:omp_fluct} OMP profiles of (a) fluctuation amplitudes of $\phi_1=\phi-\langle\phi\rangle_\varphi$, $p_e$, and $p_i$, showing the more drift-wave-like turbulence with flutter. (b) $\eta_i=L_n/L_{T_i}$ and the ratio of fluctuation amplitudes $A(T_i)/A(n)$ showing the more unstable ITG mode with flutter. Comparison is made between the model without flutter (solid lines) and with flutter (dashed lines).}  
\end{figure}

We are interested in the competition between $A(T_i)$ and $A(n)$ for ion turbulence. 
The ratio $A(T_i)/A(n)$ is compared between the cases with/without flutter by the red lines in Fig. \ref{fig:omp_eta_Ti}.
We find that without flutter $A(T_i)\leq A(n)$ in the whole edge and SOL regions, suggesting the dominance of the DW over the ITG instability.
With flutter, $A(T_i)/A(n)$ is amplified everywhere resulting in $A(T_i) > A(n)$ in the edge.
% , which also agrees with the variations of the Fourier amplitudes $|T_i|$ and $|n|$ on large scale $k\rho_s<0.5$ in Fig.~\ref{fig:fourier_all_rho990}. 
This indicates that the ITG instability gains importance when flutter is present.
With a more unstable ITG, the $T_i$ fluctuations superseding $n$ fluctuations are promoted to a greater free energy source for ion heat transport. 
This agrees with our previous observation of Fig.~\ref{fig:zonal_flux_ExB_Ti} that the ion $E \times B$ heat conduction gains relative significance against advection after both being reduced by flutter.

We find that flutter destabilizes ITG with the help of the increased $\eta_i$. 
ITG is generally known to be triggered by higher values of $\eta_i = L_n/L_{Ti} = \partial_r \log T_i/\partial_r \log n$, where $T_i$ and $n$ are the background profiles. 
The OMP profiles of $\eta_i$ are depicted by the black lines in Fig. \ref{fig:omp_eta_Ti}. 
It is noticed that $\eta_i$ increases and reaches a maximum of $1.5$ when flutter is included, while $\eta_i\leq1$ without flutter. 
Furthermore, Fig. \ref{fig:omp_eta_Ti} shows that the ratio $A(T_i)/A(n)$ perfectly follows the response of the $\eta_i$ to the inclusion of flutter.
This confirms that the higher $T_i$ fluctuation with flutter is actually related to the $\eta_i$-caused ITG instability.
It is worth noting that destabilized ITG is not necessarily accompanied by a large phase shift $\alpha_{T_i,\phi}\sim\pi/2$, as ITG has a slab branch without the curvature effect participating \citep{scott2021turbulence,manz2018microscopic}. 
% It is also noticed that $\eta_i$ remains consistently low in the SOL, which agrees with the insignificance of ITG on SOL \citep{mosetto2015finite}.

The question of why flutter increases $\eta_i$, rather than decreasing it, is important.
Fig. \ref{fig:zonal_flux_n} and Fig. \ref{fig:zonal_flux_Ti} show roughly the same reduction in particle flux and ion heat flux. 
And the similarly reduced fluxes should have led to the similarly steeper profiles of $n$ and $T_i$, while, in fact, this steepening effect differs between $n$ and $T_i$ profiles. The reason is the presence of neutral gas.
The ionization of neutrals in the SOL can inwards feed plasma density in the edge region, compensating for the effect of flutter-reduced particle flux. 
As a result, the density profile does not become as steep as the ion temperature profile does in the edge, which is evident in Fig. \ref{fig:omp_avg_n} and Fig. \ref{fig:omp_avg_ti}.
That is why $\eta_i$ becomes higher. 
The destabilization of ITG by neutrals recycling is a rich topic in itself, and has been documented in recent literature \citep{stotler2017neutral, zholobenko2021role}.
% In fact, these two runs with/without flutter share the same neutral density at the divertor.
% But the case with flutter have a slightly higher recycling rate $S_{iz}/S_{total}$

% The destabilizing effect of flutter on ITG does not contradict the commonly known electromagnetic stabilization of ITG \citep{hirose2000finite,citrin2014electromagnetic}. The so-called electromagnetic stabilization of ITG is a local effect that occurs at high beta with the electron ballooning parameter reaching a certain threshold.
% However, in our simulations, we find that flutter destabilizes ITG with the help of $\eta_i$, which is increased via a global effect. 
% For a toroidal geometry, the $\eta_i$ threshold of unstable ITG is ambiguous and depends on collisionality, electron adiabaticity, and other factors. 
% However, a larger $\eta_i$ is qualitatively conducive to destabilizing ITG as a result of including flutter.
% To demonstrate this correlation, on the same figure, we also plot the ratio of $A(T_i)$ to $A(n)$.
% It is identified that the ratio $A(T_i)/A(n)$ perfectly follows the variation of the $\eta_i$ response to the inclusion of flutter. 
% This further confirms our hypothesis that the higher $T_i$ fluctuation with flutter is actually related to the $\eta_i$ influence on ITG instability.

As discussed, flutter has a non-trivial influence on edge turbulence by stabilizing DW but destabilizing ITG.
However, we should always remember that flutter's primary effect is to stabilize the edge turbulence and reduce the transport levels, as demonstrated by Fig. \ref{fig:tracing_P} and \ref{fig:zonal_flux}.
In this context, the main role of flutter lies in stabilizing the drift-wave instability and subsequently reducing the corresponding transport.
The destabilization of ITG is a secondary consequence of the reduced DW transport and the changed $\eta_i$ profile due to the global effect and the interaction with neutral gas. 
Therefore, comprehending the impact of flutter on DW instability is the key to fully explaining the series of changes in edge turbulence dynamics induced by flutter.

\subsection{Linear analytical theory of drift-Alfv\'en waves with finite temperatures}\label{sec:linear}
In fully developed edge turbulence, linear features often become obscured \citep{scott2005drift}. 
Nevertheless, the roles of flutter in linear drift-Alfv\'en waves (DAW) are consistent with our turbulence simulations. The DAW can be seen as the finite-beta extension of drift waves, and its linear theory without temperature effects has been thoroughly discussed by \citet{scott2021turbulence}.
Under the linear approximation, flutter is manifested solely as the perpendicular gradient of electron pressure $n^{-1}\mathbf{b}_1\cdot\nabla p_e$ in Ohm's law (Eqn. \eqref{eqn:6-field:apar}).
In this study, we adhere to the classical assumption that flutter exclusively affects Ohm's law. 
Nonetheless, we extend the classical linear DAW model to incorporate finite temperature effects, aiming to explain the synchronous response of density $n$, ion temperature $T_i$, and electron temperature $T_e$ to flutter. 
With a non-zero $T_e$, the consideration of the electron thermal force $0.71 \nabla_\parallel T_e$ \citep{zeiler1997electron} becomes necessary, and its flutter counterpart $0.71 \nabla_1 T_e$ will emerge in Ohm's law.

The model is derived from the continuity equation \eqref{eqn:6-field:cont}, vorticity equation \eqref{eqn:6-field:vort}, Ohm's law \eqref{eqn:6-field:apar}, and two temperature equations \eqref{eqn:6-field:te} and \eqref{eqn:6-field:ti} based on the following assumptions.
\begin{enumerate}
    \item Background homogeneous magnetic field in a slab geometry without curvature effect.
    \item The roughly same background gradients of density and temperature profiles represented by $\omega_*\sim L_\parallel/L_n \sim L_\parallel/L_{Te}\sim L_\parallel/L_{Ti}$. The gradients are assumed to be purely radial. The $\eta_i$ and $\eta_e$ modes are excluded.    
    \item Neglected $u_\parallel\sim0$. This assumption is justified by two considerations: (1) the frequency range of the shear Alfv\'en wave is much faster than sound wave frequency under $\beta \ll 1$. (2) the $E \times B$ drift wave will also preempt the sound wave physics under the flute mode ordering $\delta \gg 1$ \citep{scott2021turbulence}
    \item The parallel heat conduction is not considered.
\end{enumerate}
\subsubsection{Formulation}
In a slab geometry, we use $z$ to represent the toroidal coordinate, $y$ for poloidal and $x$ for radial. We abbreviate $\theta=0.71$.
The $E \times B$ advection driver (from the background) is
\begin{equation}
    \mathbf{v}_E \cdot \nabla \log n
    =\mathbf{v}_E \cdot \nabla \log T_e
    =\mathbf{v}_E \cdot \nabla \log T_i
    =\omega_*\frac{\partial\phi}{\partial y}\,,
\end{equation}
and the flutter drivers for the pressure force and thermal forces in Ohm's law are
\begin{equation}
    \mathbf{b}_1 \cdot \nabla \log p_e =-2\beta_0\omega_*\frac{\partial A_\parallel}{\partial y}, \quad
    \theta \mathbf{b}_1 \cdot \nabla \log T_e =-\theta\beta_0\omega_*\frac{\partial A_\parallel}{\partial y}\,.
\end{equation}
The final `$\delta f$' linear DAW model with temperature effects reads
\begin{equation}\label{eqn:3field_linear:apar}
    \beta_0 \frac{\partial \tilde{A}_\parallel}{\partial t} 
    + \mu \frac{\partial \tilde{j}_\parallel}{\partial t}
    =
    \frac{\partial }{\partial z}\left( \tilde{n}+(1+\theta)\tilde{T}_e
    -\tilde{\phi} \right)
    - \beta_0(2+\theta)\omega_*\frac{\partial \tilde{A}_\parallel}{\partial y}
    -\eta_{\parallel 0} \tilde{j}_\parallel \,,
\end{equation}
\begin{equation}\label{eqn:3field_linear:vort}
    \frac{\partial}{\partial t} \nabla_\perp^2 \tilde{\phi} = 
    \frac{\partial \tilde{j_\parallel}}{\partial z}\,, 
\end{equation}
\begin{equation}\label{eqn:3field_linear:cont}
    \frac{\partial}{\partial t}{\tilde{n}} =-\omega_*\frac{\partial \tilde{\phi}}{\partial y}+ \frac{\partial \tilde{j_\parallel}}{\partial z}\,,
\end{equation}
\begin{equation}\label{eqn:3field_linear:Te}
    \frac{3}{2}\frac{\partial}{\partial t}{\tilde{T}_e} =-\frac{3}{2}\omega_*\frac{\partial \tilde{\phi}}{\partial y}+
    (1+\theta)\frac{\partial \tilde{j_\parallel}}{\partial z},
\end{equation}
\begin{equation}\label{eqn:3field_linear:Ti}
    \frac{3}{2}\frac{\partial}{\partial t}{\tilde{T}_i} =-\frac{3}{2}\omega_*\frac{\partial \tilde{\phi}}{\partial y}+
   \frac{\partial \tilde{j_\parallel}}{\partial z}\,,
\end{equation}
which are derived from Ohm' law \eqref{eqn:6-field:apar}, vorticity equation \eqref{eqn:6-field:vort}, continuity \eqref{eqn:6-field:cont} equation, and two temperature equations \eqref{eqn:6-field:te} and \eqref{eqn:6-field:ti}, respectively. The tilde symbol represents the fluctuation parts.
And they are closed by Ampere's law
\begin{equation}
    \tilde{j_\parallel} = -\nabla_\perp^2 \tilde{A_\parallel}\,.
\end{equation}
In Fourier space, we investigate the wave number vector
\begin{equation}
    \mathbf{k}=(k_x L_\perp,k_y L_\perp,k_z L_\parallel)/L_\parallel=(0,k_y\rho_{s},k_z)\,.
\end{equation}
And we denote the normalized wave number as $k_y\rho_{s}\equiv K$, and the frequency as $\omega = \omega_R+\gamma \mathrm{i}$ with a real growth rate $\gamma$ and a phase velocity $\omega_R$.
Then the equations above can be Fourier-transformed to
\begin{equation}\label{eqn:3field_fourier:apar}
    \left[-\omega (\beta_0+\mu K^2)
    +\beta_0(2+\theta)\omega_* K
    -i\eta_{\parallel 0} K^2\right]\tilde{j}_\parallel(\mathbf{k})
    = k_z K^2 (\tilde{n}(\mathbf{k})+(1+\theta)\tilde{T}_e(\mathbf{k})-\tilde{\phi}(\mathbf{k}))\,,
\end{equation}
\begin{equation}\label{eqn:3field_fourier:vort}
    \omega K^2\tilde{\phi}(\mathbf{k})=k_z\tilde{j}_\parallel(\mathbf{k})\,,
\end{equation}
\begin{equation}\label{eqn:3field_fourier:cont}
    \omega{\tilde{n}(\mathbf{k})}
    = \omega_* K \tilde{\phi}(\mathbf{k}) - k_z\tilde{j}_\parallel(\mathbf{k})\,,
\end{equation}
\begin{equation}\label{eqn:3field_fourier:Te}
    \frac{3}{2}\omega{\tilde{T}_e(\mathbf{k})}
    = \frac{3}{2}\omega_* K \tilde{\phi}(\mathbf{k}) - (1+\theta)k_z\tilde{j}_\parallel(\mathbf{k})\,,
\end{equation}
\begin{equation}\label{eqn:3field_fourier:Ti}
    \frac{3}{2}\omega{\tilde{T}_i(\mathbf{k})}
    = \frac{3}{2}\omega_* K \tilde{\phi}(\mathbf{k}) - k_z\tilde{j}_\parallel(\mathbf{k})\,.
\end{equation}
% We allow small phase shifts among $\tilde{n}$, $\tilde{T}_e$, and $\tilde{T}_i$
% \begin{equation}
%     \tilde{n}(\mathbf{k})/\tilde{\phi}(\mathbf{k}) \approx 1 -i \alpha_{{n},{\phi}}
% \end{equation}
% \begin{equation}
%     \frac{3}{2}\tilde{T}_e(\mathbf{k})/\tilde{n}(\mathbf{k}) \approx 1 -i \alpha_{{e},{n}}
% \end{equation}
% \begin{equation}
%     \frac{3}{2}\tilde{T}_i(\mathbf{k})/\tilde{n}(\mathbf{k}) \approx 1 -i \alpha_{{i},{n}}
% \end{equation}
%\subsubsection{transfer mechanism $\tilde{n}\rightarrow\tilde{T}_{e,i}$}
Notice that Eqns.\eqref{eqn:3field_fourier:cont}-\eqref{eqn:3field_fourier:Ti} have a similar structure.
After replacing $\tilde{j}_\parallel(\mathbf{k})$ with $\omega K^2\tilde{\phi}(\mathbf{k})/k_z$ we obtain
\begin{equation}\label{eqn:3field_fourier:n_phi}
    \frac{{\tilde{n}(\mathbf{k})}}{{\tilde{\phi}(\mathbf{k})}}
    = \frac{\omega_* K}{\omega}-K^2,
\end{equation}
\begin{equation}\label{eqn:3field_fourier:Te_phi}
    \frac{{\tilde{T}_e(\mathbf{k})}}{{\tilde{\phi}(\mathbf{k})}}
    = \frac{\omega_* K}{\omega}-\frac{2}{3}(1+\theta)K^2,
\end{equation}
\begin{equation}\label{eqn:3field_fourier:Ti_phi}
    \frac{{\tilde{T}_i(\mathbf{k})}}{{\tilde{\phi}(\mathbf{k})}}
    = \frac{\omega_* K}{\omega}-\frac{2}{3}K^2.
\end{equation}
    % \frac{{\tilde{T}_i(\mathbf{k})}}{{\tilde{n}(\mathbf{k})}}
    % = \frac{\omega_i-\frac{2}{3}\omega K}
    % {\omega_*-\omega K}
The phase shift can be approximated as $\alpha_{A,B}\approx \tan\alpha_{A,B}= \mathrm{Im}(A/B) / \mathrm{Re}(A/B) $. 
Notice that Eqn.\eqref{eqn:3field_fourier:n_phi}, \eqref{eqn:3field_fourier:Te_phi} and \eqref{eqn:3field_fourier:Ti_phi} have the same imaginary part. Hence, the magnitudes of $\alpha_{n,\phi}$, $\alpha_{T_e,\phi}$, $\alpha_{T_i,\phi}$ are proportional to the inverse real part of Eqn.\eqref{eqn:3field_fourier:n_phi}, \eqref{eqn:3field_fourier:Te_phi} and \eqref{eqn:3field_fourier:Ti_phi}, respectively.
This finding implies that the ratios of the phase shifts must be in this shape:
\begin{equation}\label{eqn:3field_fourier:e_n}
    \frac{\alpha_{T_e,\phi}}{\alpha_{n,\phi}}
    =  \left(1-\epsilon\right)/
    \left({1-\frac{2(1+\theta)}{3}\epsilon}\right),
\end{equation}
\begin{equation}\label{eqn:3field_fourier:i_n}
    \frac{\alpha_{T_i,\phi}}{\alpha_{n,\phi}}
    = \left(1-\epsilon\right)/
    \left({1-\frac{2}{3}}\epsilon\right),
\end{equation}
in which the new parameter is
\begin{equation}\label{eqn:3field_fourier:ep}
    \epsilon=K|\omega|^2/(\omega_* \omega_R) \sim \omega_R/\omega_*\,.
\end{equation}
$\epsilon$ estimates the ratio of the drift-Alfv\'en frequency $\omega_R$ to the diamagnetic frequency $\omega_*$.

Whether $\epsilon$ is a large number or a small number will lead to the distinct reduction of the model.
The accurate calculation of $\epsilon$ is crucial and will be performed after solving the dispersion relationship, which is a later-stage task.
However, a preliminary estimation indicates that $\epsilon$ must be less than 1. This estimation is based on the following reasoning.
An unstable drift wave is expected to exhibit a slightly positive phase relationship between density and potential $\alpha_{n,\phi}$ and a positive growth rate $\gamma$. 
In order to satisfy both the two conditions, Eqn. \eqref{eqn:3field_fourier:n_phi} must have a positive real part, resulting in the inequality ${\omega_* \omega_R K}/|\omega|-K^2>0$, which leads to $\epsilon<1$.

With $\epsilon<1$ and $\theta=0.71$, Eqns.~\eqref{eqn:3field_fourier:Te_phi} and \eqref{eqn:3field_fourier:Ti_phi} can be expanded as
\begin{equation}\label{eqn:an_over_ae}
    \frac{\alpha_{T_e,\phi}}{\alpha_{n,\phi}}
    = 1+\frac{2\theta-1}{3}\epsilon+O(\epsilon^2)\,,
\end{equation}
\begin{equation}\label{eqn:an_over_ai}
    \frac{\alpha_{T_i,\phi}}{\alpha_{n,\phi}}
    = 1-\frac{1}{3}\epsilon+O(\epsilon^2)\,.
\end{equation}
When $\epsilon$ is small (the justification will be checked posteriorly), $\tilde{T}_e$ can be decoupled from the phase relation between $\tilde{n}$ and $\tilde{\phi}$ in Eqn.\eqref{eqn:3field_fourier:apar}, and we obtain:
\begin{equation}
    \frac{\tilde{n}(\mathbf{k})}{\tilde{\phi}(\mathbf{k})}
    =\frac{1+\left[-\omega (\beta_0+\mu K^2)
    +\beta_0(2+\theta)\omega_* K
    -i\eta_{\parallel 0} K^2\right]\omega/k_z^2}
    {2+\theta}\,,
\end{equation}
Under the assumption of $\beta_0\ll1$ and $\mu\ll1$, the phase shift is obtained as:
\begin{equation}\label{eqn:3field_fourier:alpha}
    \alpha_{n,\phi} =[
    \underbrace{2\gamma \beta_0 \omega_R}_\text{induction}
    \underbrace{-\gamma \beta_0 (2+\theta)\omega_* K}_\text{flutter}
    \underbrace{+2 \gamma  \mu  \omega_R K^2 }_\text{inertial}
    \underbrace{+\omega_R \eta_{\parallel 0} K^2 }_\text{resistivity}]/k_z^2\,.
\end{equation}
Note that $\omega = \omega_R+\gamma i$ is not a free parameter but determined by the dispersion relation for a given $K$
% \begin{equation}
%     {\omega+\left[-\omega (\beta_0+\mu K^2)
%     +\beta_0(2+\theta)\omega_* K
%     -i\eta_{\parallel 0} K^2\right]\omega^2/k_z^2}=(2+\theta)\omega_* K-K^2(2+\theta)\omega
% \end{equation}
\begin{equation}\label{eqn:disper}
    \frac{(\beta_0+\mu K^2)\omega^3
    +[i\eta_{\parallel 0} K^2-\beta_0(2+\theta)\omega_* K]\omega^2}{(2+\theta)k_z^2}- (\frac{1}{2+\theta}+K^2)\omega + \omega_* K =0\,.
\end{equation}
It is also noticed that Eqn.~\eqref{eqn:3field_fourier:n_phi} implies the relationship between the growth rate and the phase shift
\begin{equation}\label{eqn:gamma-alpha}
    \gamma=\frac{|\omega|^2}{\omega_* K}\alpha_{n,\phi}\,.
\end{equation}
\subsubsection{Application of the analytical theory}
The analytical theory reveals the key roles of flutter in the system. 
Eqn.~\eqref{eqn:gamma-alpha} shows that only a positive phase shift can result in an unstable wave with a positive growth rate $\gamma$. Therefore, for an unstable wave with $\gamma>0$:
(1) flutter decreases the phase shift, as seen in Eqn.\eqref{eqn:3field_fourier:alpha}. In contrast, induction, inertial, and resistivity effects tend to increase the phase shift.
(2) By reducing the phase shift, flutter also decreases the growth rate, as per Eqn.~\eqref{eqn:gamma-alpha}, and hence stabilizes the wave.

\begin{figure}[!htbp]
\centering
\begin{subfigure}[b]{0.328\textwidth}
    \includegraphics[width=\textwidth]{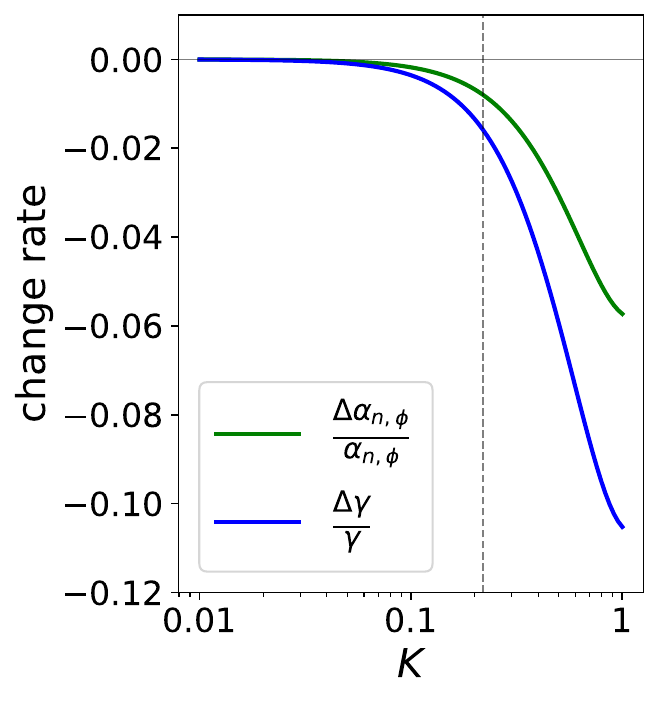}
    \caption{\label{fig:ana_delta}}
\end{subfigure}
\begin{subfigure}[b]{0.328\textwidth}
    \includegraphics[width=\textwidth]{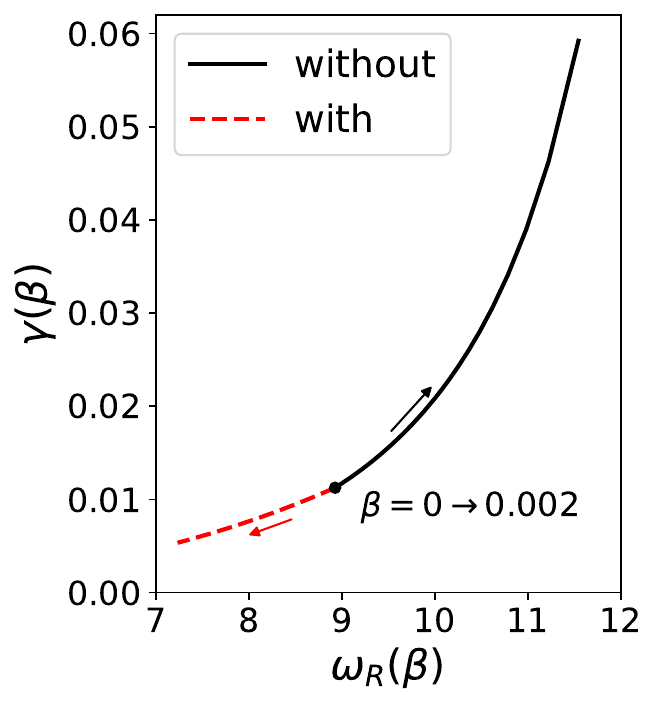}
    \caption{\label{fig:ana_root}}
\end{subfigure}
\begin{subfigure}[b]{0.32\textwidth}
    \includegraphics[width=\textwidth]{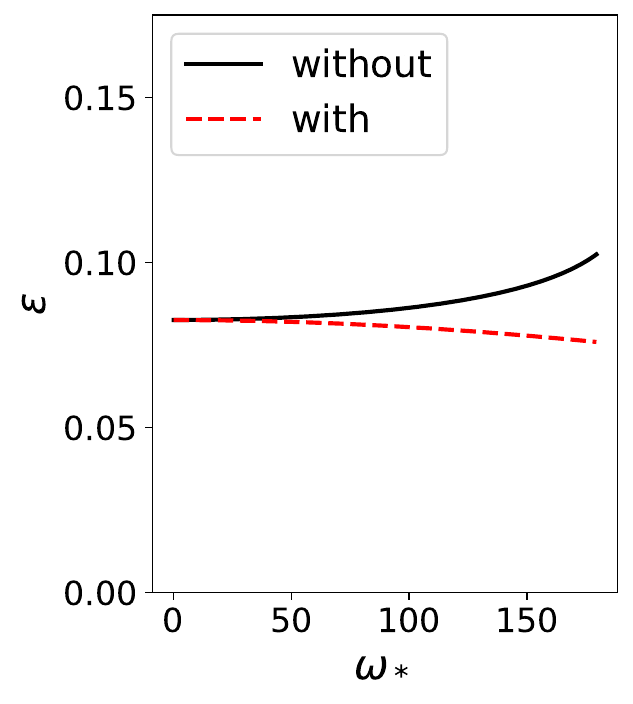}
    \caption{\label{fig:ana_epsilon}}
\end{subfigure}
\caption{\label{fig:ana_3} (a) Change rates of $\alpha_{n,\phi}$ and $\gamma$ by flutter normalised to the values without flutter, as a function of $K=k_y\rho_{s}$. Negative values indicate a decrease by flutter. The vertical dashed line corresponds to $k_y\rho_s=0.22$ marked in Fig.~\ref{fig:fourier_all_rho990}.
(b) The root locus for real frequency $\omega_R$ as x-axis and growth rate $\gamma$ as y-axis, with the increasing $\beta$ from 0 (marked by the back point) to 0.002 (the end of locus). 
(c) $\epsilon$ as a function of $\omega_*$, estimating the ratio of $\alpha_{T_e,\phi}/\alpha_{n,\phi}\sim\alpha_{T_i,\phi}/\alpha_{n,\phi}\sim 1 + O(\epsilon)$. Except for the scanned x-axis parameters, the default input parameters are $\beta=3.2\times10^{-5}$, $\mu=2.723\times 10^{-4}$, $\eta_{\parallel 0}=1.7\times 10^{-3}$ in accordance with the numerical simulations. 
$K=0.3$ and $k_z=1$ are taken.
$\omega_*=32$ is measured from the simulation results with flutter at $\rho=0.995$ of OMP.}  
\end{figure}

The variations of $\alpha_{n,\phi}$ and $\gamma$ induced by flutter are presented in Fig.~\ref{fig:ana_delta}. Notably, flutter impacts both $\alpha_{n,\phi}$ and $\gamma$ concurrently but becomes significant only at approximately $k\rho_s>0.22$, as indicated by the vertical dashed lines.
(1) Regarding the decreased phase shift, this analytical theory qualitatively explains our simulation results in Fig.~\ref{fig:fourier_all_rho990} (bottom).
(2) The linear growth rate of micro-instabilities typically correlates positively with the fluctuation amplitudes in saturated turbulence. 
Thus, the reduced $\gamma$ predicted by this analytical theory aligns well with the decreased fluctuation amplitudes of $n$ and $\phi$ displayed in Fig.~\ref{fig:fourier_all_rho990} (middle).
(3) Concerning the effective wave-number range, the analytical theory yields a reasonable range of $>0.22$, consistent with the simulation results in Fig.~\ref{fig:fourier_all_rho990}. This range results from the competition among flutter, induction, inertial, and resistivity in Eqn.\eqref{eqn:3field_fourier:alpha}.
At smaller wave numbers, flutter is scaled by $K$, and thus, it is outperformed by induction. 
However, at higher wave numbers, the effective range of flutter remains broad and is not overshadowed by inertia and resistivity. 
This occurs due to the indirect effect of flutter on the root $\omega_R$ and $\gamma$, which scale the parts of phase shift lifted by inertia and resistivity.
Nevertheless, it is worth noting that the linear model fails in the quantitative agreement with the simulations, e.g., $\alpha_{n,\phi}$ drops by less than 10\% in the linear model, but by up to 50\% in the simulations.

The impact of flutter becomes significantly accentuated as $\beta$ increases. Fig.~\ref{fig:ana_root} illustrates two root loci, one with flutter and the other without, as $\beta$ increases from 0 to 0.02. 
At the black point where $\beta=0$, the two loci coincide as flutter is zero.
As $\beta$ increases, the inclusion of flutter leads to a vastly different scenario compared to the case without flutter. Without flutter, the growth rate surges dramatically, while with flutter, it remains low and even decreases. 
This observation indicates that higher $\beta$ values strengthen the stabilizing effect of flutter.
Consequently, including flutter in simulations becomes even more crucial in higher beta scenarios, such as high confinement modes, as it plays a pivotal role in tempering the growth rate and promoting stability.

It is important to understand why the phase shift of temperatures $\alpha_{T_e,\phi}$ and $\alpha_{T_i,\phi}$ also decreases akin to $\alpha_{n,\phi}$
% , as it is the major mechanism for the reduced ExB heat conduction.
% This phenomenon can also be understood by this analytical theory. 
In Fig.~\ref{fig:ana_epsilon}, we plot $\epsilon=K|\omega|^2/(\omega_* \omega_R)$ as a function of $\omega_*$ and observe the following points:
(1) The system satisfies $\epsilon<0.1$ quite well when $\omega_*$ varies in a wide range.
% (2) $\epsilon$ is bounded instead of $\epsilon\rightarrow \infty$ as $\omega_*\rightarrow0$ because $\omega$ is not fixed but decreased by $\omega_*$. 
% This means that even if the ExB advection, which is scaled by $\omega_*$, is very small, the drift-Alfv\'en response must be even slower to allow an unstable wave.
(2) The inclusion of flutter lowers $\epsilon$, especially for large $\omega_*$. This can be understood by flutter slowing down $\omega_R$ and stabilizing $\gamma$ as shown in Fig.~\ref{fig:ana_root}.
Based on the above observations, we argue that the behavior of temperature phase shifts can be explained by the small $\epsilon$. 
Mathematically, it is comprehensible that $\alpha_{T_e,\phi}\sim\alpha_{T_i,\phi}\sim\alpha_{n,\phi}$ are bounded together under a small $\epsilon$ according to Eqn.~\eqref{eqn:an_over_ae} and \eqref{eqn:an_over_ai}. 
Physically, the small $\epsilon$ signifies that the frequency range $\omega$ where flutter plays a role in drift-Alfv\'en waves is much slower than the $E \times B$ advection $\omega_*$. Consequently, any slow changes in $\alpha_{n,\phi}$ caused by flutter can almost instantaneously be transferred to $\alpha_{T_i,\phi}$ and $\alpha_{T_e,\phi}$ by the rapid $E \times B$ advection. 
As a result, $\alpha_{T_i,\phi}$ and $\alpha_{T_e,\phi}$ consistently follow the decreased $\alpha_{n,\phi}$ by flutter.

Lastly, the role of the electron thermal force is also noticed. 
Firstly, in the phase shift \eqref{eqn:3field_fourier:alpha}, the thermal force reinforces the effect of flutter by a factor of $2.71/2$ without introducing any qualitative changes.
Secondly, in Eqn.~\eqref{eqn:an_over_ae}, the thermal force modifies the phase shift ratio, resulting in $\alpha_{T_e,\phi}$ being slightly larger than $\alpha_{n,\phi}$. This qualitative relationship of $\alpha_{T_e,\phi}>\alpha_{n,\phi}$ is also observed in our simulations, as depicted in Fig.~\ref{fig:fourier_all_rho990} (bottom). 
Moreover, the difference between $\alpha_{T_e,\phi}$ and $\alpha_{n,\phi}$ is expected to decrease with flutter due to the smaller $\epsilon$ as predicted by the linear model. 
This trend also agrees with our simulations.
However, quantitatively, the difference between $\alpha_{T_e,\phi}$ and $\alpha_{n,\phi}$ in simulations is larger than the prediction of the linear theory. 
It appears that the thermal force is one of the reasons behind $\alpha_{T_e,\phi}>\alpha_{n,\phi}$. 
There may be other factors missed in this linear model, e.g. the parallel heat conduction \citep{hallatschek2000nonlocal}.

\subsection{Numerical experiments} \label{sec:numexp}
\subsubsection{Nonlinear stabilization}
\begin{figure}[!htbp]
\centering
    \includegraphics[width=0.5\textwidth]{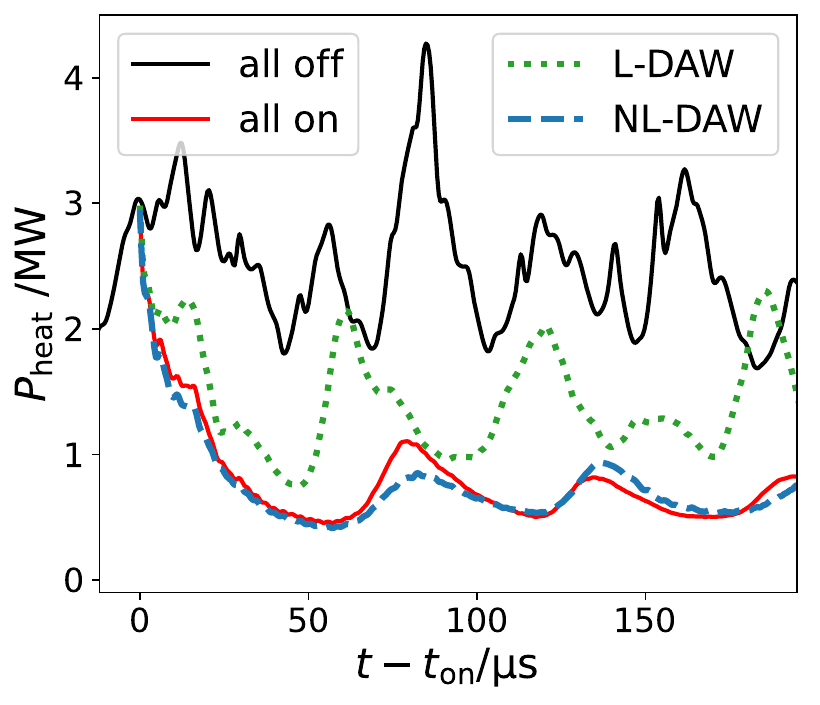}
\caption{\label{fig:test_P} Numerical experiments testing flutter terms at work, comparing the time tracing plots for the total heating power injection $P_\mathrm{heat}$.
At $t_\mathrm{on}=2.38\mathrm{ms}$ of the saturated run without flutter (labeled by `all off'), the various subsets of flutter terms are abruptly switched on.
`L-DAW': the linearized flutter terms in the DAW system, namely $\mathbf{b_1}\cdot\nabla \bar{n}$, $\mathbf{b_1}\cdot\nabla\bar{T_e}$ and $\mathbf{b_1}\cdot\nabla\bar{\phi}$ in Ohm's law \eqref{eqn:6-field:apar} with the overbar donating fixed background fields.
`NL-DAW': the necessary nonlinear flutter terms in the DAW system, namely $\mathbf{b_1}\cdot\nabla n$, $\mathbf{b_1}\cdot\nabla{T_e}$ and $\mathbf{b_1}\cdot\nabla{\phi}$ in Ohm's law \eqref{eqn:6-field:apar} and $\nabla \cdot {j_\parallel \mathbf{b_1}}$ in both the continuity equation \eqref{eqn:6-field:cont} and vorticity equation \eqref{eqn:6-field:vort}.
`all on': all flutter terms in the whole Braginskii equation set \eqref{eqn:6-field:cont}-\eqref{eqn:6-field:G}.
% The descending $P_\mathrm{heat}$ signals the stabilization by flutter.
% In the inset, the organ zone measures the discrepancy between L-DAW and NL-DAW, showing that the linear stabilization is dominating within $1 \mathrm{\mu s}$, beyond which the nonlinear stabilization takes effect.
}
\end{figure}

The linear analytical DAW theory successfully addresses the critical question of the reduced phase shifts and stabilized drift waves, providing a qualitatively explanation.
Nevertheless, it is important to check whether the nonlinear flutter contributions are stabilizing or destabilizing.
For this purpose, we performed the following numerical experiments with GRILLIX.
The results are shown in Fig.~\ref{fig:test_P}.
Starting from the saturated run without flutter (`all off'),
we switch on some flutter terms and track how the heating power $P_\mathrm{heat}$ goes.
(1) We switch on all flutter terms throughout the whole Braginskii equation set (`all on', red solid line).
The power injection immediately goes down as expected.
(2) We switch on only the linear flutter terms in the DAW model (`L-DAW' by the green dotted line).
This subset only considers the linearized interplay between the changing magnetic perturbation $\mathbf{b}_1$ and the fixed background fields, namely 
$\mathbf{b_1}\cdot\nabla \bar{n}$, $\mathbf{b_1}\cdot\nabla\bar{T_e}$ and $\mathbf{b_1}\cdot\nabla\bar{\phi}$ in Ohm's law \eqref{eqn:6-field:apar}.
It can be seen that L-DAW initiates the descending power injection, which, however, always remains higher than the case `all on'. 
% In addition, we note a finding which is not visualized in this plot: the stabilization mainly originates from $\mathbf{b_1}\cdot\nabla \bar{n}$ and $\mathbf{b_1}\cdot\nabla\bar{T_e}$, while the contribution from $\mathbf{b_1}\cdot\nabla\bar{\phi}$ is marginal.  
(3) We switch on some nonlinear flutter  terms in the DAW system.
Although the flutter nonlinearity occurs all over the system of equations, in every parallel operator,  
we identify that only four nonlinear flutter terms are necessary by the method of trial and error.
This minimal subset (`NL-DAW', blue dashed line) contains $\mathbf{b_1}\cdot\nabla n$, $\mathbf{b_1}\cdot\nabla{T_e}$ and $\mathbf{b_1}\cdot\nabla{\phi}$ in Ohm's law \eqref{eqn:6-field:apar} and $\nabla \cdot {j_\parallel \mathbf{b_1}}$ in both the continuity equation \eqref{eqn:6-field:cont} and vorticity equation \eqref{eqn:6-field:vort}.
With the flutter terms in NL-DAW, the heating power can almost exactly recover the trajectory of the case `all on'.

Based on the above observations, we reach two important conclusions.
Firstly, flutter enters the edge turbulence mainly via the nonlinear DAW system.
Despite flutter's presence in every parallel operator in the whole equation set, only the minimal subset NL-DAW demonstrates a noteworthy impact on turbulent transport. 
This finding further justifies the relevance of DAW theory, which has pointed us in the right direction.

Secondly, the nonlinear stabilization is noticeable comparing L-DAW and NL-DAW.
% We also zoom in on the first $7$ micro-seconds and plot the deviations caused by flutter nonlinearity in the inset.
The heating power of L-DAW coincides with NL-DAW only before \SI{1}{\micro s}, during which the linear stabilization is dominating. 
Afterwards, the nonlinear effect quickly comes into effect and adds to further stabilization.
In the long run, the heating power $P_\mathrm{heat}$ with NL-DAW reaches a quasi-saturation (but slowly grows), which is roughly half of $P_\mathrm{heat}$ with L-DAW.
The role of linear flutter terms has been well understood through the theory developed in the section \ref{sec:linear}, and explains our observations qualitatively well.
The flutter nonlinearity is missed in that linear theory, however, yet it turns out to be essential quantitatively in our numerical experiment.
% On top of `L-DAW', the enabled flutter thermal force $0.71\nabla_1 T_e$ in Ohm's law (`L-DAW-et') is conducive to the descending power injection to a lower level.
% This enhancement by thermal force has already been predicted by the linear theory.
% But the difference between `L-DAW-et' and `all' still remains.
% (4) On top of `L-DAW-et', we switch on the purely nonlinear flutter terms $\nabla_1 \phi$ in Ohm's law and $\nabla_1 j_\parallel$ in continuity and vorticity equation (`NL-DAW-et'), which are ignored by the linear DAW model.
% This NL-DAW-et is found to be the minimal subset of effective flutter terms.
% With NL-DAW-et, the heating power can almost recover the trajectory of the case with all terms.

% While the DAW theory offers insights into the behavior observed in the simulations, it is important to acknowledge that the DAW's adequacy does not necessarily imply the its indispensability. 
% In other words, it is still conceivable that flutter may play a role in non-DAW terms, such as gyro-viscosity or parallel heat conduction, leading to similar outcomes given the complexity of edge turbulence.

\subsubsection{Beta scan}
The linear theory predicts more remarkable stabilization by flutter when $\beta$ increases, as shown by Fig.~\ref{fig:ana_root}, which will be tested in this section.
Our reference simulations are based on the L-mode discharge with $\beta_0=3.227\times 10^{-5}$.
In the following numerical experiments, we double and quadruple the dimensionless parameter $\beta_0$ in the saturated reference simulation.
The changed $\beta_0$ will enter the system through the magnetic induction term $\partial A_\parallel / \partial t$ in Ohm's law \eqref{eqn:6-field:apar}, as well as all flutter terms.

\begin{figure}[!htbp]
\centering
\begin{subfigure}[b]{0.4\textwidth}
    \includegraphics[width=\textwidth]{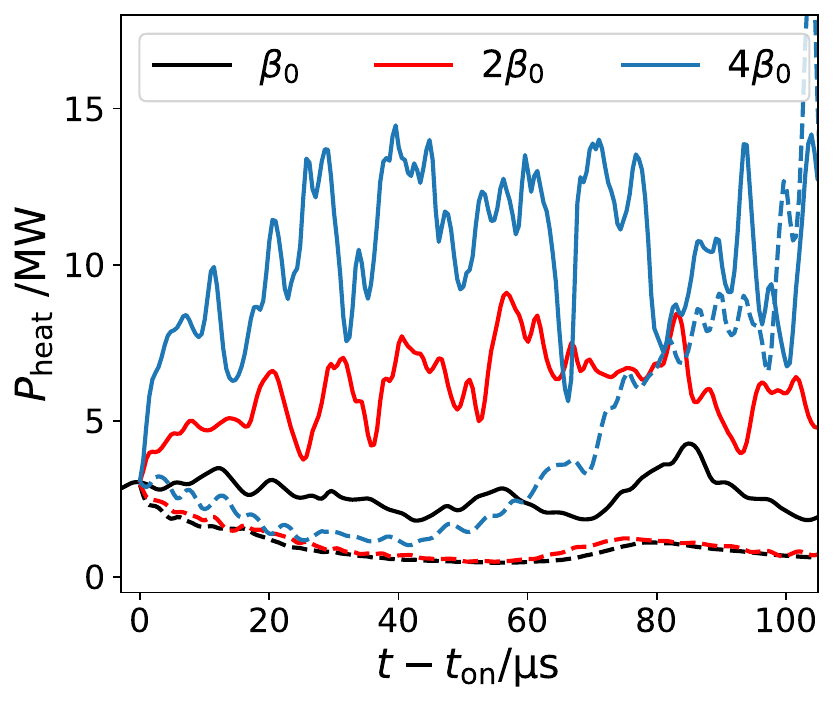}
    \caption{\label{fig:test_P_beta}}
\end{subfigure}
\begin{subfigure}[b]{0.42\textwidth}
    \includegraphics[width=\textwidth]{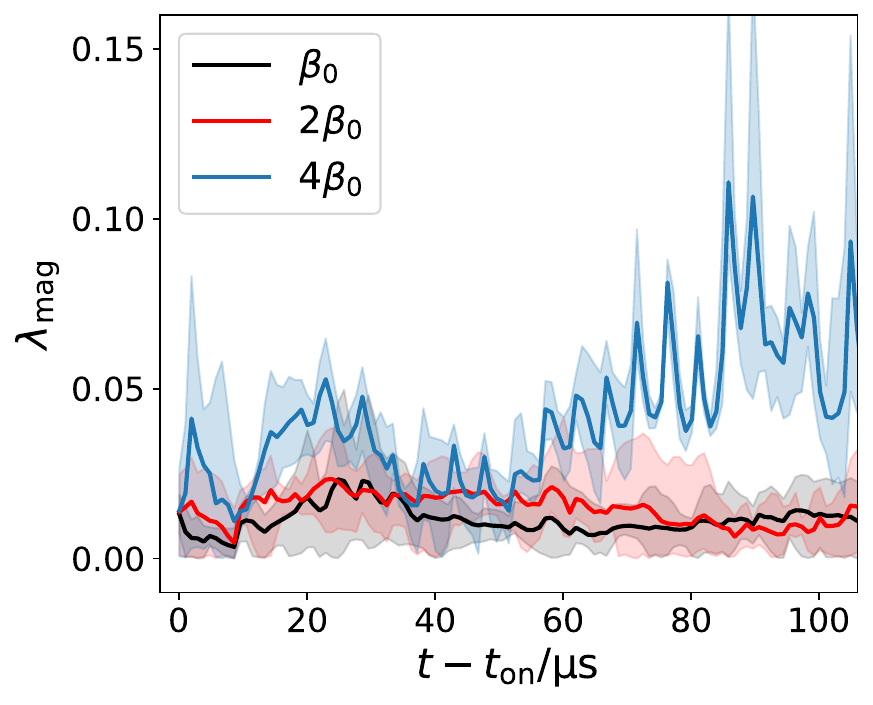}
    \caption{\label{fig:test_mag_beta}}
\end{subfigure}
\caption{\label{fig:test} Numerical experiments scanning $\beta_0$, $2\beta_0$, and $4\beta_0$. 
At $t_\mathrm{on}=\SI{2.38}{ms}$ of the saturated run without flutter, $\beta_0$ are abruptly doubled or quadrupled.
(a) The time tracing for the total heating power injection $P_\mathrm{heat}$, comparing the cases without flutter (solid) and with flutter (dashed).
(b) The time tracing for $\lambda_\mathrm{mag}=|\oiint \mathbf{Q}^\mathrm{mag}_e \cdot \mathrm{d}\mathbf{s}| /|\oiint \mathbf{Q}^\mathrm{ExB}_e \cdot \mathrm{d}\mathbf{s}|$ in the cases with flutter. The colored zones show the variation range of $\lambda_\mathrm{mag}$ when $\rho$ varies from $0.98$ to $0.995$, with the solid lines being the averages.
% The stabilizing effect of flutter is consistent in the cases of $\beta_0$ and $2\beta_0$, while qualitative changes in the flutter transport occur from $2\beta_0$ to $4\beta_0$.
}
\end{figure}

The time tracking of the heating power $P_\mathrm{heat}$ is compared among these cases of $\beta_0$, $2\beta_0$, and $4\beta_0$ with and without flutter in Fig.~\ref{fig:test_P_beta}.
Let us first examine the $2\beta_0$ case. 
Without flutter, the heating power grows immediately, suggesting the destabilizing effect of the magnetic induction.
With flutter, the heating power still decreases.
As a result of both, the $2\beta_0$ case exhibits the more pounced stabilizing effect of flutter compared to the reference $\beta_0$ case.
It is also noticed that the cases of $\beta_0$ and $2\beta_0$ exhibit very similar heating power figures when flutter is present.
In this sense, the turbulent transport level is not sensitive to $\beta$, as was also noted in the beta scan conducted by \citet{giacomin2022turbulent}.
This could be explained by Fig. \ref{fig:ana_root}, in which the linear growth rate remains flat when flutter is included.
Despite the insensitivity to $\beta$, it is not accurate to perceive electromagnetic effects as marginal.
On the contrary, both magnetic induction and flutter have individual impacts.
In terms of the overall turbulent transport level, the stabilization caused by flutter seems to counteract the destabilization of magnetic induction, but neither of the two effects is negligible.
In terms of the micro mode structure of turbulence, it is essential to consider both magnetic induction and flutter to accurately simulate the phase shift across different wave numbers, as indicated by Eqn. \eqref{eqn:3field_fourier:alpha}.

The case with $4\beta_0$ turns out to be more complicated.
When flutter is included, the heating power goes down during the first \SI{40}{\micro s} showing the familiar stabilizing effect.
Afterwards, however, the heating power with flutter increases rapidly and even tends to overtake the level without flutter.
We find that this growing heating power is accompanied by the dramatic enhancement of the magnetic flutter transport.
To measure the relative strength of the flutter transport compared to the $E \times B$ transport, we define $\lambda_\mathrm{mag}=|\oiint \mathbf{Q}^\mathrm{mag}_e \cdot \mathrm{d}\mathbf{s}| /|\oiint \mathbf{Q}^\mathrm{ExB}_e \cdot \mathrm{d}\mathbf{s}|$ as a ratio of the flutter electron heat flux to the $E \times B$ electron heat flux in terms of surface integrals.
% As a function of time and flux surfaces, $\lambda_\mathrm{mag}$ measures the relative strength of the flutter transport compared to the ExB transport on the fly.
The time tracking of $\lambda_\mathrm{mag}$ is depicted by Fig. \ref{fig:test_mag_beta}.
We can see that the cases $\beta_0$ and $2\beta_0$ have a similar $\lambda_\mathrm{mag} \sim 10^{-2}$; there is no qualitative change from $\beta_0$ to $2\beta_0$.
From $2\beta_0$ to $4\beta_0$, a qualitative change occurs, with a jump in $\lambda_\mathrm{mag}$ of more than 10.
This might be related to triggering of MHD instabilities with flutter, which will be studied in the future.

\section*{Conclusion}
In this study, we conducted comprehensive global simulations for edge turbulence in ASDEX Upgrade at full scale using GRILLIX, which now incorporates magnetic flutter effects. A comparative analysis was performed between simulations with and without flutter inclusion.
Even in low-$\beta$ L-mode discharges, the incorporation of flutter has a significant impact, reducing turbulent transport levels by a factor of 2. 
This reduction brings the heating power closer to experimental values. 
While the direct transport caused by flutter remains significantly smaller than the $E \times B$ transport by two orders of magnitude, including flutter leads to a remarkable reduction of about 50\% in the $E \times B$ transport of particles, electron thermal energy, and ion thermal energy.
The consequence of decreased turbulent transport is evident in the altered OMP profiles of plasma density, electron temperature, and radial electric field. Notably, the profiles align with experimental observations more closely when flutter effects are considered.
Our study strongly affirms that the inherent dynamics of edge turbulence consistently reside within electromagnetic regimes even under L-mode conditions. Furthermore, we emphasize the essential role of flutter in predictive simulations of reactor-relevant edge turbulence.

The key mechanism for reduced $E \times B$ transport is found to be flutter decreasing the phase shifts $\alpha_{n,\phi}$, $\alpha_{T_e,\phi}$ and $\alpha_{T_i,\phi}$, leading to stabilized drift-wave turbulence.
To explain the overall reduction in all three phase shifts, we have developed the linear analytical theory for DAW instability with finite temperature effect, which is an extension of the classical linear DAW model by \citet{scott2001low}.
The flutter gradient force $\mathbf{b_1}\cdot\nabla p_e$ in Ohm's law is found to reduce $\alpha_{n,\phi}$, while the uniform $E \times B$ advection terms in continuity and temperature equations tend to homogenize $\alpha_{n,\phi}\sim\alpha_{T_e,\phi}\sim\alpha_{T_i,\phi}$.
The frequency of an unstable drift-Alfv\'en wave with flutter at work is always much slower than the frequency of $E \times B$ advection.
Therefore, the effect of flutter on $\alpha_{n,\phi}$ can be effectively passed down to $\alpha_{T_e,\phi}$ and $\alpha_{T_i,\phi}$, which in turn are reduced by flutter as well.
The presence of flutter electron thermal force in Ohm's law can linearly reinforce the phase shift reduction.

Despite the success of this linear DAW theory, the dominance of the stabilization by linear flutter terms is merely confined within very short time scale of \SI{1}{\micro s}.
Over a longer time scale, the flutter nonlinearity will add substantially to the stabilization.
Although every flutter term is essentially nonlinear,  
the necessary nonlinearity for stabilization only lies in  
$\mathbf{b_1}\cdot\nabla n$, $\mathbf{b_1}\cdot\nabla{T_e}$ and $\mathbf{b_1}\cdot\nabla{\phi}$ in Ohm's law \eqref{eqn:6-field:apar}, and $\nabla \cdot {j_\parallel \mathbf{b_1}}$ in both continuity equation \eqref{eqn:6-field:cont} and vorticity equation \eqref{eqn:6-field:vort}.
We note that the electromagnetic nonlinear stabilization of ITG turbulence has been well studied for the core region \citep{citrin2014electromagnetic,whelan2018nonlinear}.
Our finding, on the other hand, encourages exploring the electromagnetic nonlinear stabilization of DAW turbulence in the edge region in the future.

As a secondary result of the stabilized DAW, the reduced $E \times B$ flux tends to steepen density and temperature profiles, by which flutter increases $\eta_i$ and destabilizes ITG. 
The destabilized ITG, however, only slightly modifies the overall reduced transport, as it leads to a minor effect on ion $E \times B$ heat flux, making it more conductive with higher $T_i$ fluctuations.
%More accurately predicting $\eta_i$ will require the recycling neutral boundary condition, which will be available in GRILLIX in the future.

Lastly, the $\beta$ scan shows that the stabilization by flutter could be even stronger with increasing $\beta$, until the flutter transport becomes noticeable. 
Considering both the 50\% $E \times B$ transport reduction by flutter in low $\beta$ scenarios and the enhanced flutter transport in high $\beta$ scenarios, magnetic flutter may play a crucial role in shaping the transport behavior during the L-H transition, which will be studied in the future. 
% This present work paves the way for studying L-H transition with GRILLIX in the near future.

% In conclusion, this study proves that flutter is essential for predictive reactor-relevant edge turbulence simulations even in low beta scenarios. 
% This present work also paves the way for simulating high confinement modes and L-H transition with GRILLIX in the future.

\section*{Acknowledgments}
The authors thank Jan Pfennig and Christoph Pitzal for helpful discussions. 
This work has been carried out within the framework of the EUROfusion Consortium, funded by the European Union via the Euratom Research and Training Programme (Grant Agreement No.101052200 — EUROfusion). 
Views and opinions expressed are however those of the author(s) only and do not necessarily reflect those of the European Union or the European Commission. 
Neither the European Union nor the European Commission can be held responsible for them.

\section*{Appendix}\label{appe}
The typical result of simulations with different free parameters is listed in Table. \ref{tab:other}.
As those free parameters change, the turbulence regimes and heating power also greatly change.
However, the role of flutter is always consistent: 
(1) to make $A(\phi_1)/A(p_e)$ closer to 1 (more drift-wave-like), and
(2) to decrease $P_\mathrm{heat}$ (stabilization).
Moreover, as the turbulence becomes more and more drift-wave-like from case 1 to case 3, the stabilization effect of flutter also becomes stronger.
This parallel comparison again confirms our notion that the magnetic flutter decreases the edge transport by stabilizing the drift-wave turbulence in reactor-relevant simulations. 
\begin{table}[!ht]
\centering
\begin{tabular}{|c|c|c|c|c|}
\hline
                  & flutter & 
{\makecell{\textbf{case 1} (ref.)\\
\scriptsize{$\alpha^\mathrm{FS}_e=0.1, \alpha^\mathrm{FS}_i=1.0$}\\
\scriptsize{$N_\mathrm{div}=0.02, T_{i}^\mathrm{core}=2$}}} &
{\makecell{case 2\\
\scriptsize{$\alpha^\mathrm{FS}_e=0.3, \alpha^\mathrm{FS}_i=0.3$}\\
\scriptsize{$N_\mathrm{div}=0.04, T_{i}^\mathrm{core}=3$}}} & {\makecell{case 3\\
\scriptsize{$\alpha^\mathrm{FS}_e=1.0, \alpha^\mathrm{FS}_i=1.0$}\\
\scriptsize{$N^\mathrm{div}=0.02, T_{i}^\mathrm{core}=2$}}}            \\
\hline
\multirow{2}{*}{\makecell{$A(\phi_1)/A(p_e)$}} 
& without  & {$1.542$} &  {$1.540$} &  {$1.174$} \\ 
&  with  & {$1.159$} &   {$0.915$} &   {$1.025$}     \\ \hline
\multirow{2}{*}{\makecell{$P_\mathrm{heat}/\mathrm{MW}$ }} & without& {$2.59_{\pm0.54}$}& {$2.37_{\pm0.40}$}  &  {$0.485_{\pm0.03}$} \\ 
 & with & {$1.34_{\pm0.30}$} & {$1.05_{\pm0.04}$}  &                  {$0.120_{\pm0.001}$}  \\ \hline 
& \makecell{stabilization\\factor} & \textbf{1.93} & \textbf{2.25} &                  \textbf{4.04}  \\
 \hline
\end{tabular}
\caption{\label{tab:other} Summary of simulations with varied free parameters.
Case 1 is the reference.
$A(\phi_1)/A(p_e)$ is the ratio of the fluctuation amplitudes of $\phi_1$ to $p_e$ corresponding to Fig.~\ref{fig:omp_fluct_phi_pei}.
$A(\phi_1)/A(p_e)\sim1$ indicates more drift-wave-like turbulence.
$P_\mathrm{heat}$ is the total heating power corresponding to Fig.~\ref{fig:tracing_P}.
The stabilization factor is calculated as the ratio of $P_\mathrm{heat}$ without flutter to $P_\mathrm{heat}$ with flutter. 
}
\end{table}

\bibliography{main.bib}% Syntax for version >= 1.2

@book{scott2021turbulence,
  title={Turbulence and Instabilities in Magnetised Plasmas, Volume 1: Fluid drift turbulence},
  author={Scott, Bruce},
  year={2021},
  publisher={IOP Publishing}
}

@article{zeiler1997nonlinear,
  title={Nonlinear reduced Braginskii equations with ion thermal dynamics in toroidal plasma},
  author={Zeiler, A and Drake, JF and Rogers, B},
  journal={Physics of Plasmas},
  volume={4},
  number={6},
  pages={2134--2138},
  year={1997},
  publisher={American Institute of Physics}
}

@book{scott2021turbulence2,
  title={Turbulence and Instabilities in Magnetised Plasmas, Volume 2: Gyrokinetic theory and gyrofluid turbulence},
  author={Scott, Bruce},
  year={2021},
  publisher={IOP Publishing}
}

@article{naulin2003electromagnetic,
  title={Electromagnetic transport components and sheared flows in drift-Alfven turbulence},
  author={Naulin, Volker},
  journal={Physics of Plasmas},
  volume={10},
  number={10},
  pages={4016--4028},
  year={2003},
  publisher={American Institute of Physics}
}

@article{scott1997three,
  title={Three-dimensional computation of drift {Alfv{\'e}n} turbulence},
  author={Scott, Bruce},
  journal={Plasma Physics and Controlled Fusion},
  volume={39},
  number={10},
  pages={1635},
  year={1997},
  publisher={IOP Publishing}
}

@article{DANNERT200467,
title = {Vlasov simulation of kinetic shear Alfvén waves},
journal = {Computer Physics Communications},
volume = {163},
number = {2},
pages = {67-78},
year = {2004},
issn = {0010-4655},
doi = {https://doi.org/10.1016/j.cpc.2004.09.001},
author = {Tilman Dannert and Frank Jenko}
}

@article{Dudson_2021,
doi = {10.1088/1361-6587/ac2af9},
url = {https://dx.doi.org/10.1088/1361-6587/ac2af9},
year = {2021},
month = {oct},
publisher = {IOP Publishing},
volume = {63},
number = {12},
pages = {125008},
author = {B D Dudson and S L Newton and J T Omotani and J Birch},
title = {On Ohm’s law in reduced plasma fluid models},
journal = {Plasma Physics and Controlled Fusion}
}

@article{doerk2011gyrokinetic,
  title={Gyrokinetic microtearing turbulence},
  author={Doerk, H and Jenko, F and Pueschel, MJ and Hatch, DR},
  journal={Physical Review Letters},
  volume={106},
  number={15},
  pages={155003},
  year={2011},
  publisher={APS}
}

@article{hirshman1978neoclassical,
  title={Neoclassical current in a toroidally-confined multispecies plasma},
  author={Hirshman, Steven P},
  journal={The Physics of Fluids},
  volume={21},
  number={8},
  pages={1295--1301},
  year={1978},
  publisher={American Institute of Physics}
}

@phdthesis{scott2001low,
  Type = {habilitation},
  title={Low frequency fluid drift turbulence in magnetised plasmas},
  author={Scott, Bruce},
  school={Düsseldorf University},
  year={2001},
  institution={Max-Planck-Institut f{\"u}r Plasmaphysik}
}

@phdthesis{manz2018microscopic,
  Type = {habilitation},
  title={The microscopic picture of plasma edge turbulence},
  author={Manz, Peter},
  year={2018},
  school={Technische Universit{\"a}t M{\"u}nchen}
}

@article{pueschel2010transport,
  title={Transport properties of finite-$\beta$ microturbulence},
  author={Pueschel, MJ and Jenko, F},
  journal={Physics of Plasmas},
  volume={17},
  number={6},
  pages={062307},
  year={2010},
  publisher={American Institute of Physics}
}

@article{hirose2000finite,
  title={On finite $\beta$ stabilization of the toroidal ion temperature gradient mode},
  author={Hirose, A},
  journal={Physics of Plasmas},
  volume={7},
  number={2},
  pages={433--436},
  year={2000},
  publisher={American Institute of Physics}
}

@article{callen1977drift,
  title={Drift-wave turbulence effects on magnetic structure and plasma transport in tokamaks},
  author={Callen, JD},
  journal={Physical Review Letters},
  volume={39},
  number={24},
  pages={1540},
  year={1977},
  publisher={APS}
}

@article{zholobenko2021role,
  title={The role of neutral gas in validated global edge turbulence simulations},
  author={Zholobenko, W and others},
  journal={Nuclear Fusion},
  volume={61},
  number={11},
  pages={116015},
  year={2021},
  publisher={IOP Publishing}
}

@article{mandell2020electromagnetic,
  title={Electromagnetic full-gyrokinetics in the tokamak edge with discontinuous Galerkin methods},
  author={Mandell, NR and Hakim, A and Hammett, GW and Francisquez, M},
  journal={Journal of Plasma Physics},
  volume={86},
  number={1},
  pages={905860109},
  year={2020},
  publisher={Cambridge University Press}
}

@article{jenko1999numerical,
  title={Numerical computation of collisionless drift Alfv{\'e}n turbulence},
  author={Jenko, Frank and Scott, Bruce D},
  journal={Physics of Plasmas},
  volume={6},
  number={7},
  pages={2705--2713},
  year={1999},
  publisher={American Institute of Physics}
}

@article{giacomin2021theory,
  title={Theory-based scaling laws of near and far scrape-off layer widths in single-null L-mode discharges},
  author={Giacomin, Maurizio and Stagni, Adriano and Ricci, Paolo and Boedo, Jose A and Horacek, Jan and Reimerdes, Holger and Tsui, CK},
  journal={Nuclear Fusion},
  volume={61},
  number={7},
  pages={076002},
  year={2021},
  publisher={IOP Publishing}
}

@article{jenko2001nonlinear,
  title={Nonlinear electromagnetic gyrokinetic simulations of tokamak plasmas},
  author={Jenko, Frank and Dorland, W},
  journal={Plasma physics and controlled fusion},
  volume={43},
  number={12A},
  pages={A141},
  year={2001},
  publisher={IOP Publishing}
}

@article{eich2021separatrix,
  title={The separatrix operational space of ASDEX Upgrade due to interchange-drift-Alfv{\'e}n turbulence},
  author={Eich, Thomas and Manz, Peter and ASDEX Upgrade Team and others},
  journal={Nuclear Fusion},
  volume={61},
  number={8},
  pages={086017},
  year={2021},
  publisher={IOP Publishing}
}

@article{scott2007tokamak,
  title={Tokamak edge turbulence: background theory and computation},
  author={Scott, Bruce D},
  journal={Plasma Physics and Controlled Fusion},
  volume={49},
  number={7},
  pages={S25},
  year={2007},
  publisher={IOP Publishing}
}

@article{kobayashi2020physics,
  title={The physics of the mean and oscillating radial electric field in the L--H transition: the driving nature and turbulent transport suppression mechanism},
  author={Kobayashi, T},
  journal={Nuclear Fusion},
  volume={60},
  number={9},
  pages={095001},
  year={2020},
  publisher={IOP Publishing}
}

@article{zholobenko2021electric,
  title={Electric field and turbulence in global Braginskii simulations across the ASDEX Upgrade edge and scrape-off layer},
  author={Zholobenko, W and others},
  journal={Plasma Physics and Controlled Fusion},
  volume={63},
  number={3},
  pages={034001},
  year={2021},
  publisher={IOP Publishing}
}

@article{giacomin2020investigation,
  title={Investigation of turbulent transport regimes in the tokamak edge by using two-fluid simulations},
  author={Giacomin, Maurizio and Ricci, Paolo},
  journal={Journal of Plasma Physics},
  volume={86},
  number={5},
  pages={905860502},
  year={2020},
  publisher={Cambridge University Press}
}

@article{stegmeir2019global,
  title={Global turbulence simulations of the tokamak edge region with GRILLIX},
  author={Stegmeir, A and Ross, A and Body, T and Francisquez, M and Zholobenko, W and Coster, D and Maj, O and Manz, P and Jenko, F and Rogers, BN and others},
  journal={Physics of Plasmas},
  volume={26},
  number={5},
  pages={052517},
  year={2019},
  publisher={AIP Publishing LLC}
}

@article{STEGMEIR2023108801,
title = {Analysis of locally-aligned and non-aligned discretisation schemes for reactor-scale tokamak edge turbulence simulations},
journal = {Computer Physics Communications},
volume = {290},
pages = {108801},
year = {2023},
issn = {0010-4655},
doi = {https://doi.org/10.1016/j.cpc.2023.108801},
author = {A. Stegmeir and T. Body and W. Zholobenko}
}

@Article{Body2019,
  author    = {T. Body and A. Stegmeir and W. Zholobenko and D. Coster and F. Jenko},
  journal   = {Contributions to Plasma Physics},
  title     = {Treatment of advanced divertor configurations in the flux{-}coordinate independent turbulence code {GRILLIX}},
  year      = {2019},
  doi       = {doi.org/10.1002/ctpp.201900139},
  eprint    = {e201900139},
  publisher = {Wiley},
}

@Article{Zholobenko2019,
  author    = {W. Zholobenko and A. Stegmeir and T. Body and A. Ross and P. Manz and O. Maj and D. Coster and F. Jenko and M. Francisquez and B. Zhu and B.N. Rogers},
  journal   = {Contributions to Plasma Physics},
  title     = {Thermal dynamics in the flux-coordinate independent turbulence code {GRILLIX}},
  year      = {2019},
  doi       = {10.1002/ctpp.201900131},
  eprint    = {e201900131},
  publisher = {Wiley},
}

@article{tang1978microinstability,
  title={Microinstability theory in tokamaks},
  author={Tang, Wo Mo},
  journal={Nuclear Fusion},
  volume={18},
  number={8},
  pages={1089},
  year={1978},
  publisher={IOP Publishing}
}

@article{dong1987finite,
  title={Finite beta effects on ion temperature gradient driven modes},
  author={Dong, JQ and Guzdar, PN and Lee, YC},
  journal={The Physics of fluids},
  volume={30},
  number={9},
  pages={2694--2702},
  year={1987},
  publisher={American Institute of Physics}
}

@article{kim1993electromagnetic,
  title={Electromagnetic effect on the toroidal ion temperature gradient mode},
  author={Kim, JY and Horton, W and Dong, JQ},
  journal={Physics of Fluids B: Plasma Physics},
  volume={5},
  number={11},
  pages={4030--4039},
  year={1993},
  publisher={American Institute of Physics}
}

@article{whelan2018nonlinear,
  title={Nonlinear electromagnetic stabilization of plasma microturbulence},
  author={Whelan, GG and Pueschel, MJ and Terry, PW},
  journal={Physical Review Letters},
  volume={120},
  number={17},
  pages={175002},
  year={2018},
  publisher={APS}
}

@article{ritz1989fluctuation,
  title={Fluctuation-induced energy flux in the tokamak edge},
  author={Ritz, Ch P and Bravenec, RV and Schoch, PM and Bengtson, RD and Boedo, JA and Forster, JC and Gentle, KW and He, Y and Hickok, RL and Kim, YJ and others},
  journal={Physical review letters},
  volume={62},
  number={16},
  pages={1844},
  year={1989},
  publisher={APS}
}

@article{zweben2007edge,
  title={Edge turbulence measurements in toroidal fusion devices},
  author={Zweben, SJ and Boedo, JA and Grulke, O and Hidalgo, C and LaBombard, B and Maqueda, RJ and Scarin, P and Terry, JL},
  journal={Plasma Physics and Controlled Fusion},
  volume={49},
  number={7},
  pages={S1},
  year={2007},
  publisher={IOP Publishing}
}

@article{garcia2007collisionality,
  title={Collisionality dependent transport in TCV SOL plasmas},
  author={Garcia, Odd Erik and Pitts, RA and Horacek, J and Madsen, Jens and Naulin, Volker and Nielsen, Anders Henry and Rasmussen, J Juul},
  journal={Plasma Physics and Controlled Fusion},
  volume={49},
  number={12B},
  pages={B47},
  year={2007},
  publisher={IOP Publishing}
}

@article{citrin2014electromagnetic,
  title={Electromagnetic stabilization of tokamak microturbulence in a high-$\beta$ regime},
  author={Citrin, J and Garcia, J and G{\"o}rler, T and Jenko, F and Mantica, P and Told, D and Bourdelle, C and Hatch, DR and Hogeweij, GMD and Johnson, Thomas and others},
  journal={Plasma Physics and Controlled Fusion},
  volume={57},
  number={1},
  pages={014032},
  year={2014},
  publisher={IOP Publishing}
}

@article{snyder2001electromagnetic,
  title={Electromagnetic effects on plasma microturbulence and transport},
  author={Snyder, PB and Hammett, GW},
  journal={Physics of Plasmas},
  volume={8},
  number={3},
  pages={744--749},
  year={2001},
  publisher={American Institute of Physics}
}

@article{zhu2021drift,
  title={Drift reduced Landau fluid model for magnetized plasma turbulence simulations in BOUT++ framework},
  author={Zhu, Ben and Seto, Haruki and Xu, Xue-qiao and Yagi, Masatoshi},
  journal={Computer Physics Communications},
  volume={267},
  pages={108079},
  year={2021},
  publisher={Elsevier}
}

@article{giacomin2022gbs,
  title={The GBS code for the self-consistent simulation of plasma turbulence and kinetic neutral dynamics in the tokamak boundary},
  author={Giacomin, M and Ricci, P and Coroado, A and Fourestey, G and Galassi, D and Lanti, E and Mancini, D and Richart, N and Stenger, LN and Varini, N},
  journal={Journal of Computational Physics},
  volume={463},
  pages={111294},
  year={2022},
  publisher={Elsevier}
}

@article{hager2022electromagnetic,
  title={Electromagnetic total-f algorithm for gyrokinetic particle-in-cell simulations of boundary plasma in XGC},
  author={Hager, Robert and Ku, S and Sharma, AY and Chang, CS and Churchill, RM and Scheinberg, A},
  journal={Physics of Plasmas},
  volume={29},
  number={11},
  pages={112308},
  year={2022},
  publisher={AIP Publishing LLC}
}

@article{hager2020gyrokinetic,
  title={Gyrokinetic understanding of the edge pedestal transport driven by resonant magnetic perturbations in a realistic divertor geometry},
  author={Hager, Robert and Chang, CS and Ferraro, NM and Nazikian, R},
  journal={Physics of Plasmas},
  volume={27},
  number={6},
  pages={062301},
  year={2020},
  publisher={AIP Publishing LLC}
}

@article{chang2017gyrokinetic,
  title={Gyrokinetic projection of the divertor heat-flux width from present tokamaks to ITER},
  author={Chang, Choong Seock and Ku, S and Loarte, Alberto and Parail, Vassili and Koechl, Florian and Romanelli, Michele and Maingi, Rajesh and Ahn, J-W and Gray, T and Hughes, J and others},
  journal={Nuclear Fusion},
  volume={57},
  number={11},
  pages={116023},
  year={2017},
  publisher={IOP Publishing}
}

@article{manz2015origin,
  title={Origin and turbulence spreading of plasma blobs},
  author={Manz, P and Ribeiro, TT and Scott, BD and Birkenmeier, G and Carralero, D and Fuchert, G and M{\"u}ller, SH and M{\"u}ller, HW and Stroth, U and Wolfrum, E},
  journal={Physics of Plasmas},
  volume={22},
  number={2},
  pages={022308},
  year={2015},
  publisher={AIP Publishing LLC}
}

@article{manz2020diffusion,
  title={The diffusion limit of ballistic transport in the scrape-off layer},
  author={Manz, P and Hufnagel, C and Zito, A and Carralero, D and Griener, M and Lunt, T and Pan, O and Passoni, M and Tal, B and Wischmeier, M and others},
  journal={Physics of Plasmas},
  volume={27},
  number={2},
  pages={022506},
  year={2020},
  publisher={AIP Publishing LLC}
}

@article{scott2006edge,
  title={Edge turbulence and its interaction with the equilibrium},
  author={Scott, BD},
  journal={Contributions to Plasma Physics},
  volume={46},
  number={7-9},
  pages={714--725},
  year={2006},
  publisher={Wiley Online Library}
}

@article{scott2005drift,
  title={Drift wave versus interchange turbulence in tokamak geometry: Linear versus nonlinear mode structure},
  author={Scott, Bruce D},
  journal={Physics of Plasmas},
  volume={12},
  number={6},
  year={2005},
  publisher={AIP Publishing}
}

@article{fundamenski2005parallel,
  title={Parallel heat flux limits in the tokamak scrape-off layer},
  author={Fundamenski, W},
  journal={Plasma physics and controlled fusion},
  volume={47},
  number={11},
  pages={R163},
  year={2005},
  publisher={IOP Publishing}
}

@article{arakawa1997computational,
  title={Computational design for long-term numerical integration of the equations of fluid motion: Two-dimensional incompressible flow. Part I},
  author={Arakawa, Akio},
  journal={Journal of computational physics},
  volume={135},
  number={2},
  pages={103--114},
  year={1997},
  publisher={Elsevier}
}

@article{stegmeir2016field,
  title={The field line map approach for simulations of magnetically confined plasmas},
  author={Stegmeir, Andreas and Coster, David and Maj, Omar and Hallatschek, Klaus and Lackner, Karl},
  journal={Computer Physics Communications},
  volume={198},
  pages={139--153},
  year={2016},
  publisher={Elsevier}
}

@article{giacomin2022turbulent,
  title={Turbulent transport regimes in the tokamak boundary and operational limits},
  author={Giacomin, M and Ricci, P},
  journal={Physics of Plasmas},
  volume={29},
  number={6},
  year={2022},
  publisher={AIP Publishing}
}

@article{zholobenko2023filamentary,
  title={Filamentary transport in global edge-SOL simulations of ASDEX Upgrade},
  author={Zholobenko, W and Pfennig, J and Stegmeir, A and Body, T and Ulbl, P and Jenko, F and ASDEX Upgrade Team and others},
  journal={Nuclear Materials and Energy},
  volume={34},
  pages={101351},
  year={2023},
  publisher={Elsevier}
}

@article{mantica1991broadband,
  title={Broadband fluctuations and particle transport in the edge plasma during ECRH in DITE},
  author={Mantica, P and Vayakis, G and Hugill, J and Cirant, S and Pitts, RA and Matthews, GF},
  journal={Nuclear fusion},
  volume={31},
  number={9},
  pages={1649},
  year={1991},
  publisher={IOP Publishing}
}

@article{hallatschek2000nonlocal,
  title={Nonlocal simulation of the transition from ballooning to ion temperature gradient mode turbulence in the tokamak edge},
  author={Hallatschek, K and Zeiler, A},
  journal={Physics of Plasmas},
  volume={7},
  number={6},
  pages={2554--2564},
  year={2000},
  publisher={American Institute of Physics}
}

@article{stegmeir2017advances,
  title={Advances in the flux-coordinate independent approach},
  author={Stegmeir, Andreas and Maj, Omar and Coster, David and Lackner, Karl and Held, Markus and Wiesenberger, Matthias},
  journal={Computer Physics Communications},
  volume={213},
  pages={111--121},
  year={2017},
  publisher={Elsevier}
}

@article{bufferand2021progress,
  title={Progress in edge plasma turbulence modelling—hierarchy of models from 2D transport application to 3D fluid simulations in realistic tokamak geometry},
  author={Bufferand, H and Bucalossi, J and Ciraolo, G and Falchetto, G and Gallo, A and Ghendrih, Ph and Rivals, N and Tamain, P and Yang, H and Giorgiani, G and others},
  journal={Nuclear Fusion},
  volume={61},
  number={11},
  pages={116052},
  year={2021},
  publisher={IOP Publishing}
}

@article{paruta2019blob,
  title={Blob velocity scaling in diverted tokamaks: A comparison between theory and simulation},
  author={Paruta, Paola and Beadle, C and Ricci, P and Theiler, C},
  journal={Physics of Plasmas},
  volume={26},
  number={3},
  pages={032302},
  year={2019},
  publisher={AIP Publishing LLC}
}

@article{braginskii1965transport,
  title={Transport processes in a plasma},
  author={Braginskii, SI},
  journal={Reviews of plasma physics},
  volume={1},
  pages={205},
  year={1965}
}

@article{naulin2005shear,
  title={Shear flow generation and energetics in electromagnetic turbulence},
  author={Naulin, V and Kendl, A and Garcia, OE and Nielsen, AH and Rasmussen, J Juul},
  journal={Physics of Plasmas},
  volume={12},
  number={5},
  year={2005},
  publisher={AIP Publishing}
}

@article{zweben2002edge,
  title={Edge turbulence imaging in the Alcator C-Mod tokamak},
  author={Zweben, SJ and Stotler, DP and Terry, JL and LaBombard, B and Greenwald, M and Muterspaugh, M and Pitcher, CS and Alcator C-Mod Group and Hallatschek, K and Maqueda, RJ and others},
  journal={Physics of Plasmas},
  volume={9},
  number={5},
  pages={1981--1989},
  year={2002},
  publisher={American Institute of Physics}
}

@Article{Hirshman1981,
  author    = {S.P. Hirshman and D.J. Sigmar},
  journal   = {Nuclear Fusion},
  title     = {Neoclassical transport of impurities in tokamak plasmas},
  year      = {1981},
  month     = {sep},
  number    = {9},
  pages     = {1079--1201},
  volume    = {21},
  doi       = {10.1088/0029-5515/21/9/003},
  publisher = {{IOP} Publishing},
}

@article{Salari_MMS_2000,
    title = "{Code Verification by the Method of Manufactured Solutions}",
    author = {Kambitz Salari and Patrick Knupp},
    doi = {10.2172/759450},
    url = {https://www.osti.gov/biblio/759450},
    place = {United States},
    year = {2000}
}

@book{helander2005collisional,
  title={Collisional transport in magnetized plasmas},
  author={Helander, Per and Sigmar, Dieter J},
  volume={4},
  year={2005},
  publisher={Cambridge university press}
}

@article{zeiler1997electron,
  title={Electron temperature fluctuations in drift-resistive ballooning turbulence},
  author={Zeiler, Andreas and Drake, James F and Biskamp, Dieter},
  journal={Physics of Plasmas},
  volume={4},
  number={4},
  pages={991--1001},
  year={1997},
  publisher={American Institute of Physics}
}

@article{walkden2022physics,
  title={The physics of turbulence localised to the tokamak divertor volume},
  author={Walkden, Nicholas and Riva, Fabio and Harrison, James and Militello, Fulvio and Farley, Thomas and Omotani, John and Lipschultz, Bruce},
  journal={Communications Physics},
  volume={5},
  number={1},
  pages={139},
  year={2022},
  publisher={Nature Publishing Group UK London}
}

@article{stotler2017neutral,
  title={Neutral recycling effects on ITG turbulence},
  author={Stotler, DP and Lang, J and Chang, CS and Churchill, RM and Ku, S},
  journal={Nuclear Fusion},
  volume={57},
  number={8},
  pages={086028},
  year={2017},
  publisher={IOP Publishing}
}

@article{michels2021gene,
  title={GENE-X: A full-f gyrokinetic turbulence code based on the flux-coordinate independent approach},
  author={Michels, Dominik and Stegmeir, Andreas and Ulbl, Philipp and Jarema, Denis and Jenko, Frank},
  journal={Computer Physics Communications},
  volume={264},
  pages={107986},
  year={2021},
  publisher={Elsevier}
}

@article{hidalgo1995edge,
  title={Edge turbulence and anomalous transport in fusion plasmas},
  author={Hidalgo, C},
  journal={Plasma Physics and Controlled Fusion},
  volume={37},
  number={11A},
  pages={A53},
  year={1995},
  publisher={IOP Publishing}
}

\end{document}